\documentclass[aps,pra,twocolumn,showpacs,superscriptaddress,floatfix]			{revtex4-1}
\usepackage[colorlinks=true,citecolor=blue, linkcolor=blue, urlcolor=blue]{hyperref}

\usepackage{graphicx}
\usepackage{epstopdf}
\usepackage{amsmath}
\usepackage{amssymb}
\usepackage{slashed}
\usepackage{braket}
\usepackage{setspace}
\usepackage{mathrsfs}
\usepackage[usenames, dvipsnames]{color}
\usepackage[caption=false,position=top,justification=raggedright,singlelinecheck=off,labelformat=parens,subrefformat=parens]{subfig}
\usepackage{lipsum}

\def\Jvol<#1,#2,#3>{#1}
\def\Jpage<#1,#2,#3>{#2}
\def\Jyear<#1,#2,#3>{#3}

\begin{document}

\title{Macroscopic Quantum Tunneling Escape of Bose-Einstein Condensates}

\author{Xinxin Zhao}
\affiliation{International Center for Quantum Materials, School of Physics, Peking University, Beijing 100871, China}
\affiliation{Department of Physics, Colorado School of Mines, Golden, Colorado 80401, USA}

\author{Diego A. \surname{Alcala}}
\author{Marie A. \surname{McLain}}
\author{Kenji \surname{Maeda}}
\affiliation{Department of Physics, Colorado School of Mines, Golden, Colorado 80401, USA}

\author{Shreyas \surname{Potnis}}
\author{Ramon \surname{Ramos}}
\affiliation{Centre for Quantum Information and Quantum Control and Institute for Optical sciences,Department of Physics and Institute of Optics, University of Toronto, 60 St. George Street, Toronto, Ontario M5S 1A7, Canada}

\author{Aephraim M. \surname{Steinberg}}
\affiliation{Centre for Quantum Information and Quantum Control and Institute for Optical sciences,Department of Physics and Institute of Optics, University of Toronto, 60 St. George Street, Toronto, Ontario M5S 1A7, Canada}
\affiliation{Canadian Institute For Advanced Research, 180 Dundas Street West, Toronto, Ontario M5G 1Z8, Canada}

\author{Lincoln D. \surname{Carr}}
\affiliation{Department of Physics, Colorado School of Mines, Golden, Colorado 80401, USA}

\date{\today}

\begin{abstract}

Recent experiments on macroscopic quantum tunneling reveal a non-exponential decay of the number of atoms trapped in a quasibound state behind a potential barrier. Through both experiment and theory, we demonstrate this non-exponential decay results from interactions between atoms. Quantum tunneling of tens of thousands of $^{87}$Rb atoms in a Bose-Einstein condensate is modeled by a modified Jeffreys-Wentzel-Kramers-Brillouin model, taking into account the effective time-dependent barrier induced by the mean-field. Three-dimensional Gross-Pitaevskii simulations corroborate a mean-field result when compared with experiments. However, with one-dimensional modeling using time-evolving block decimation, we present an effective renormalized mean-field theory that suggests many-body dynamics for which a bare mean-field theory may not apply.

\end{abstract}

\pacs{}
\maketitle

\section{Introduction}

  \begin{figure}
 \subfloat[]{
 \includegraphics[width=8.0 cm]{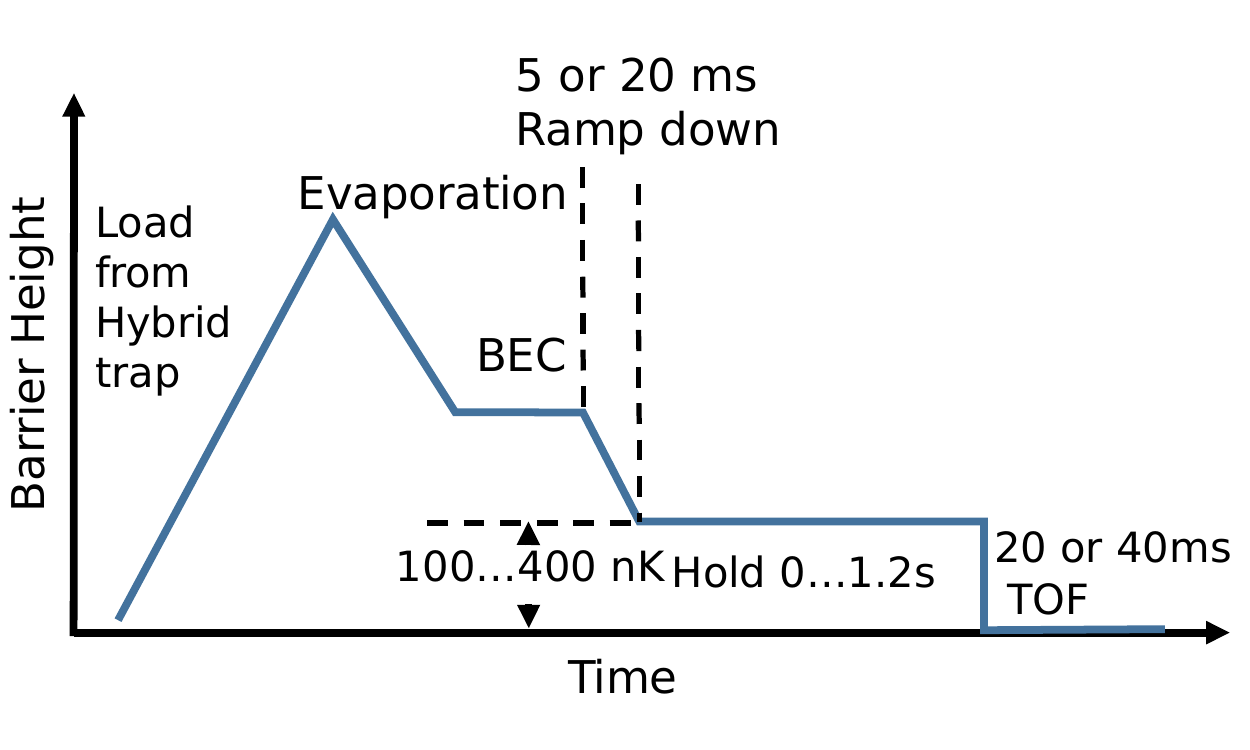}
 \label{fig:Torontoexp:a}
 }
 \hfill
 \vspace{-5mm}
 \subfloat[]{
 \includegraphics[width=8.0 cm]{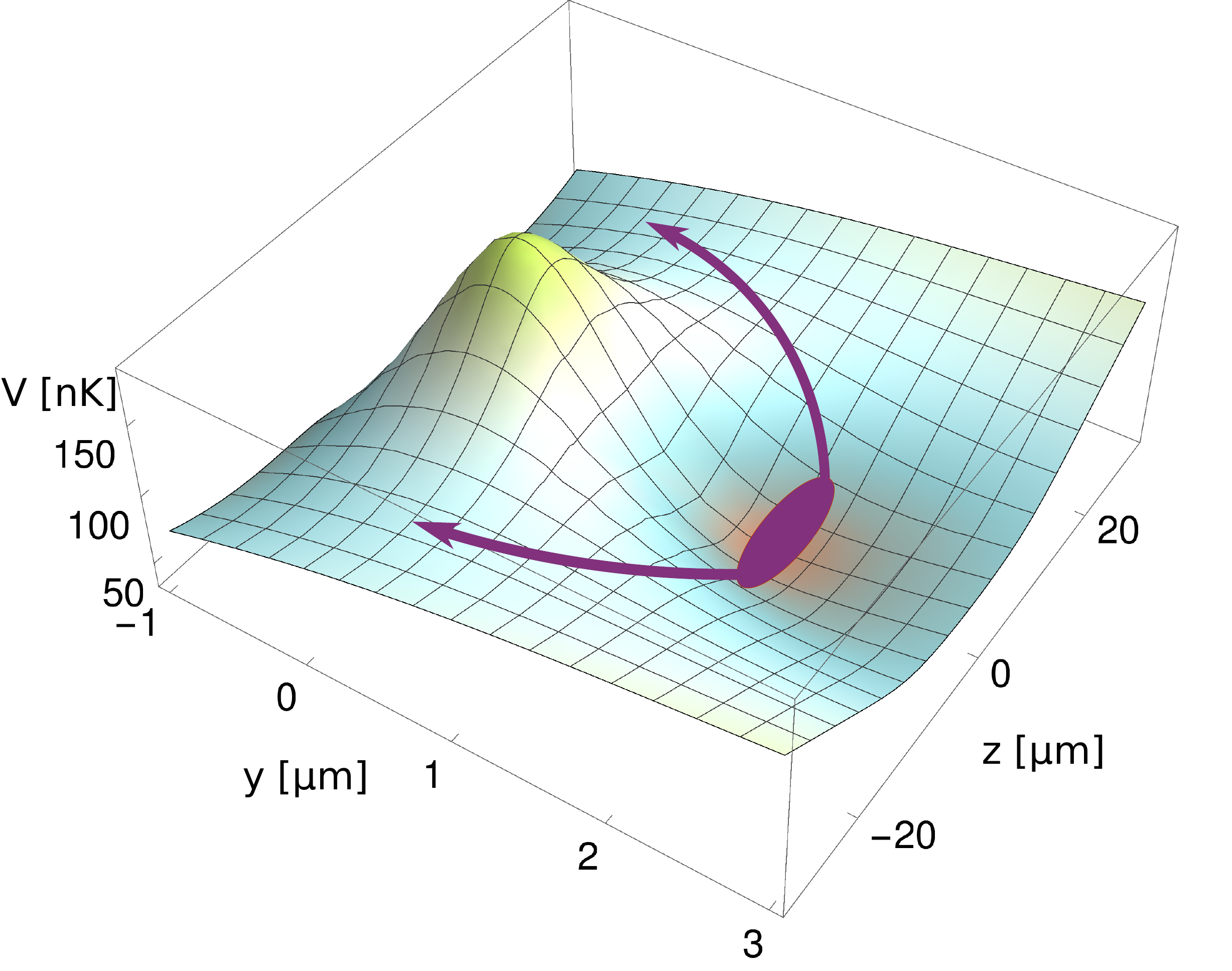}
 \label{fig:Torontoexp:b}
 }
 \hfill
 \vspace{-5mm}
 \subfloat[]{
 \includegraphics[width=8.0 cm]{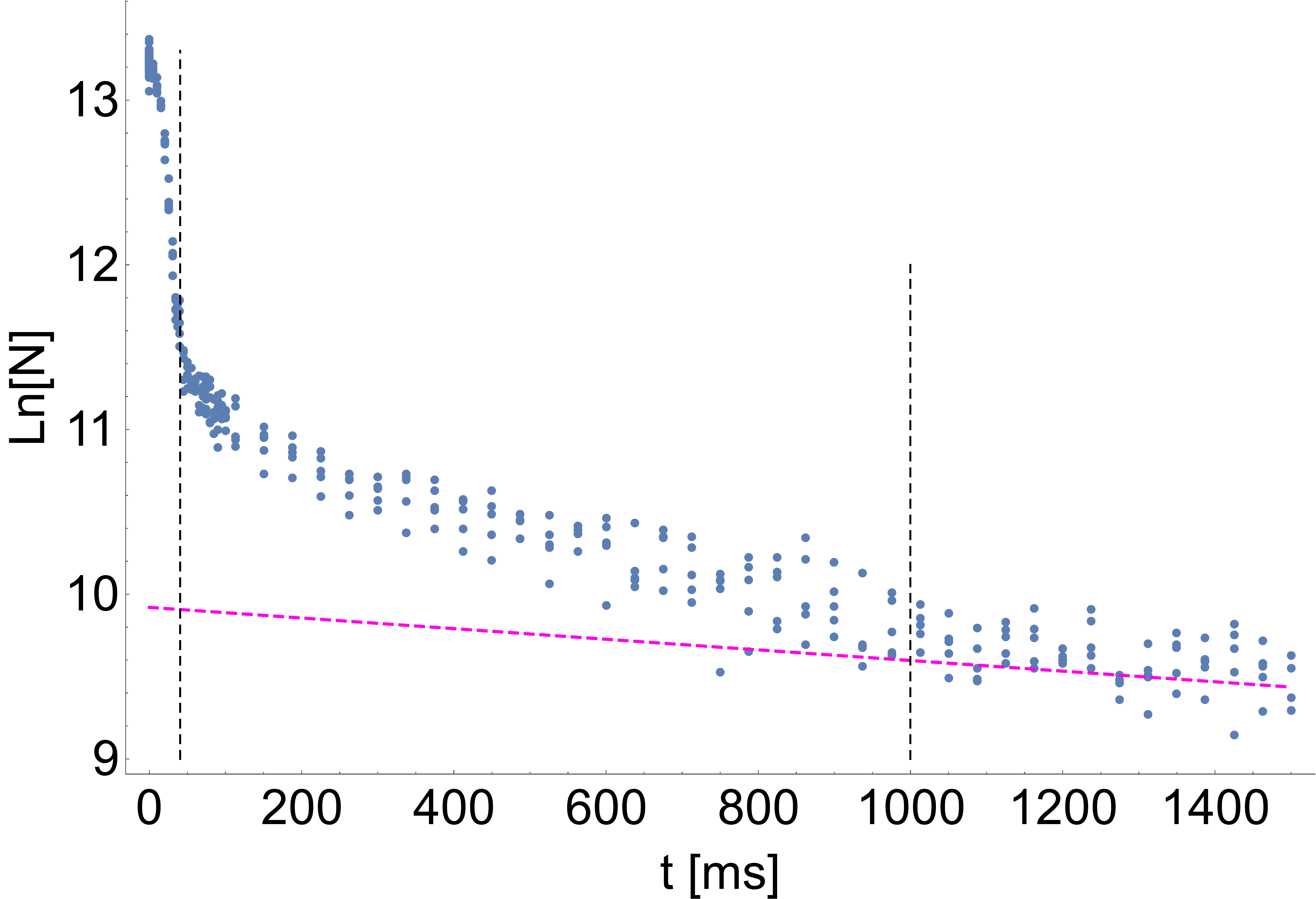}
 \label{fig:Torontoexp:c}
 }
 \caption{\textit{Macroscopic Quantum Tunneling Experiment.} (a) Schematic of experimental time sequence to obtain MQT in a BEC. (b) Experimental 3D potential with a barrier height of 190\,nK (peak height) in the weak configuration. The distance between saddle points to the potential minimum is $x_{0}=18(1) \,\mathrm{\mu m}$. Sketched is a BEC trapped in the local minimum of the potential (purple ellipse) escaping via the weakest part of the potential (purple arrows). (c) Experimental data, in number of trapped atoms $N$, exhibits a non-exponential decay, in contrast to the exponential background decay due to atomic losses (pink dashed line).  The vertical dashed lines divide the experimental dynamics into the three dominant regions of classical spilling, MQT, and background decay.\label{fig:Torontoexp}}
\end{figure}

Quantum tunneling, first studied in $\alpha$-decay, is one of the most significant and earliest effects observed in quantum mechanics~\cite{alpha1929}.
Tunneling provided a basis for the theory of molecular spectra in the 1920s, when it was demonstrated that transitions between chiral isomers occurred at measurable rates~\cite{hund1927}. This applies to biological systems, where the transition rate between isomers is slow enough to allow stable life, and even to cosmology, where the transition for polar molecules such as ammonia shows a measurable rate~\cite{pwanderson}. Tunneling has applications in many different systems, from two-proton decay and double beta decay in nuclear physics to the interdisciplinary study of tunneling in enzymes~\cite{jrommel}, with roots in biology, chemistry, and physics. As electronic devices reach the nanoscale, quantum tunneling will play a larger and larger role in understanding and developing nanoelectronics, such as tunneling diodes~\cite{tundiodes}. Moreover, with the advent of Josephson junctions, the epitome of macroscopic quantum devices based on tunneling, we can now measure voltage with unprecedented accuracy~\cite{nistjj}.

The definitions and details of macroscopic quantum tunneling (MQT), quantum tunnelling at macroscopic scales, is discussed in Section~\ref{section:statistic}. Studying MQT in the context of Bose-Einstein condensates (BECs) presents many advantages in terms of both fundamental explorations and future MQT device design. First, BECs offer a high degree of controllability: interactions can be tuned over seven orders of magnitude using Feshbach resonances~\cite{feshbach}. Second, experimental advances in radio-frequency (RF) magnetic traps~\cite{magneticTrap}, as well as optical trapping~\cite{opticalTrap}, allow for greater access to controllable experiments needed to study MQT. Third, BECs enable manipulation of many-body states~\cite{manyBodyExample} that are inaccessible in other experimental settings. Many-body simulations, which we discuss next, elucidate the importance of this point.  Fourth, BECs have controllable statistics and spatial dimensions.  Fifth, BECs allow site-resolved microscopy in the context of optical lattices in 1, 2, and 3 dimensions.  Sixth, atomic interferometry now permits observation of up to tenth order correlators for quasi-1D systems~\cite{Langen2015}.

To complement experimental advances, numerically exact simulations of bosonic Josephson junctions using the Bose-Hubbard model demonstrate a substantial deviation from mean-field theory~\cite{Sakmann2009}. However, until recently MQT has largely been treated with semiclassical approximations, such as in a double well~\cite{Albiez2005,Raghavan1999}, in Landau-Zener tunneling~\cite{Cristiani2002}, and in escape tunneling~\cite{Moiseyev2004,Carr2005,dekel2007}. Theoretical methods for these semiclassical estimations include Jeffreys-Wentzel-Kramers-Brillouin (JWKB) and instanton approaches, along with the nonlinear Schr\"odinger equation (NLS) and hydrodynamic formulations thereof. Beyond these mean-field and instanton techniques, matrix product state methods such as time-evolving block decimation (TEBD) and multi-configurational time-dependent Hartree for bosons (MCTDHB) theory both provide numerical solutions of many-body dynamics. For example, the use of TEBD to simulate the Bose-Hubbard model for superfluid decay, has confirmed numerical limits on instanton computations~\cite{Danshita2010,Danshita2012}. Non-Hermitian quantum mechanics, which is frequently used in scattering problems and now applied to tunneling problems as an effective model, is broadening the view of tunneling phenomena, such as its use in asymmetric tunneling and interchain pair tunneling~\cite{nonhermi2014, nonhermi2017}. Another possible many-body method, the time-adaptive MCTDHB, was used in our Josephson example above~\cite{Sakmann2009}. MCTDHB has also been applied to the quantum escape problem~\cite{Beinke2015,Lode2012}; though this work examined depletion, the method has not yet produced predictions for entanglement. Measures such as entropy, entanglement, and correlations help illustrate when semiclassical or lower-order mean-field approximations fail~\cite{Diego2016}.

In this Article, we present a combined theoretical and experimental study of MQT. Single-particle quantum tunneling can be modeled with the Schr\"odinger equation, with well-known exponential decay in the number of atoms trapped over time~\cite{alpha1929}. However, we have performed a macroscopic experiment exhibiting non-exponential decay of BEC tunneling from a single trapping well to unbound space~\cite{shreyas2016}. Figures~\subref*{fig:Torontoexp:a} and \subref*{fig:Torontoexp:b} sketch the experimental process and the single well trap. Figure~\subref*{fig:Torontoexp:c} displays raw data from the experiment, presenting an example of the observed non-exponential decay. The experimental data demonstrates that atomic interactions have participated in the tunneling process. Inspired by this result, we develop an alternative theoretical model in which the interactions cause the barrier to change dynamically, leading the decay to deviate from the single-particle exponential case.  Subsequently we suggest an alternate interpretation in terms of an effective or renormalized mean-field theory drawn from TEBD simulations and accounting for the effects of condensate fragmentation and depletion, which can be tested in future MQT experiments.

This Article is organized in the following manner. We offer a brief discussion covering the nuances and regimes of quantum tunneling in Section~\ref{section:regimes}, including four subtopics: statistical properties, the role of interactions,  type of trapping potential, and dimension of the system. The aim of this discussion is to briefly touch on the vast tunneling landscape and spark interest and ignite ideas for other physicists intrigued by the many open questions in MQT research.  Next, the details of the MQT  experiment are covered in Section~\ref{section:exp}, with its experimental settings, results, and a 3D mean-field simulation.  Furthermore, we present a case study analysis that exhibits excellent agreement with the experimental non-exponential decay result and further verifies the assumption of mean-field dynamics.  In Section~\ref{section:WKB} we then address the question of simpler effective 1D models. The experiment proceeds through three distinct regimes: initial transient classical spilling, quantum tunneling, and decay dominated by background loss; we model the tunneling and decay regimes, the main subject of this Article. We use a modified JWKB method in which the inter-particle interaction is taken into account via an effective mean-field interaction parameter, and we include the background loss in the tunneling rate. The model contains two fitting parameters which help to illuminate the experimental findings.  In Section~\ref{section:TEBD}, we propose an explanation for the effective mean-field used in the JWKB model.  Motivated by advances in nonlinear optics, we demonstrate how a mean-field model can effectively reproduce many-mode many-body dynamics of the quantum tunneling process of a meta-stable state.  Finally, we summarize our conclusions and future research avenues in Section~\ref{section:Conclusions}.

\section{Regimes of macroscopic quantum tunneling}\label{section:regimes}

\begin{figure}
\centering
 \subfloat[]{
 \includegraphics[width=7.8 cm]{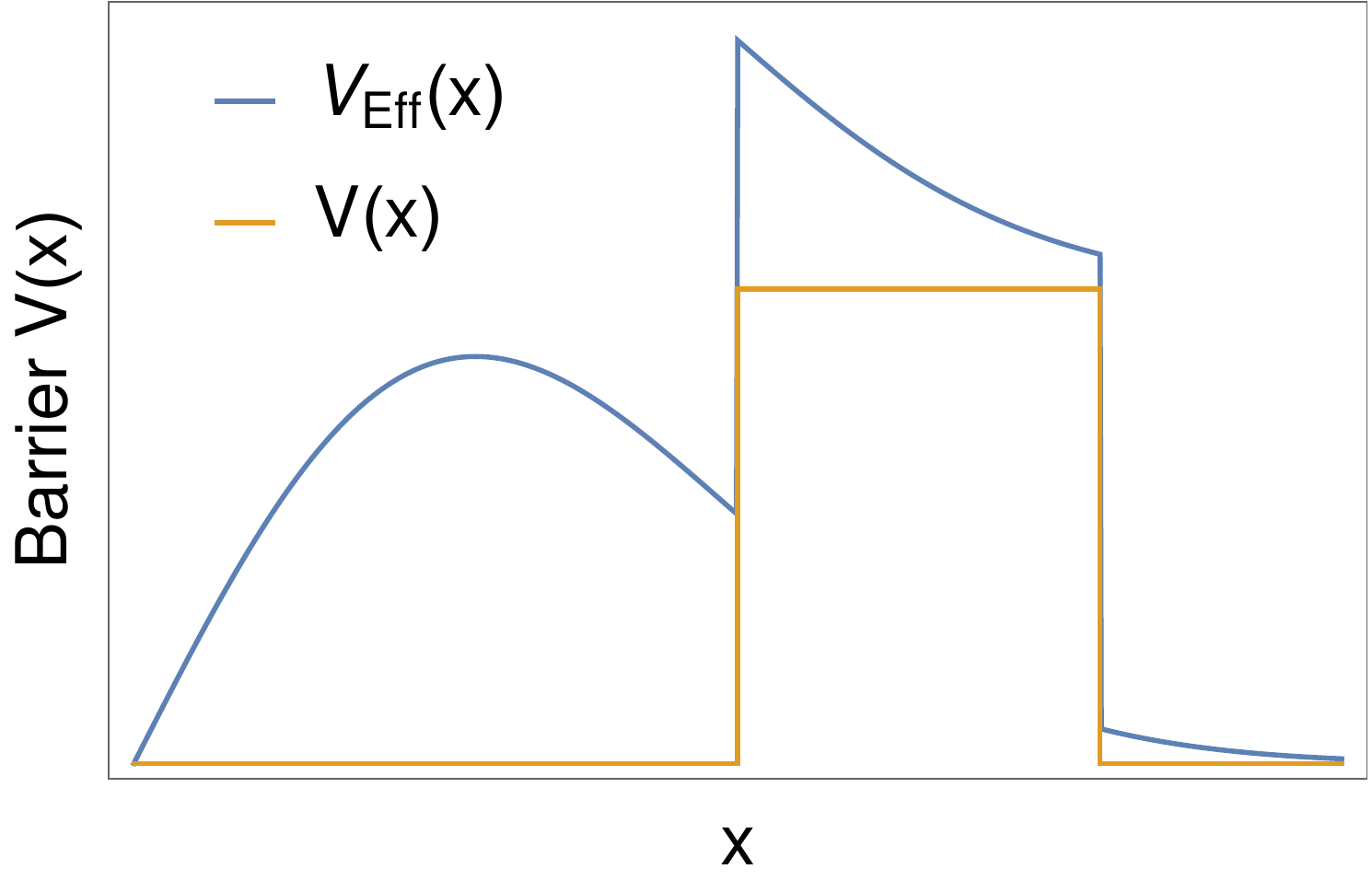}
 \label{fig:schematic:a}
 }
 \hfill
 \subfloat[]{
 \includegraphics[width=7.8 cm]{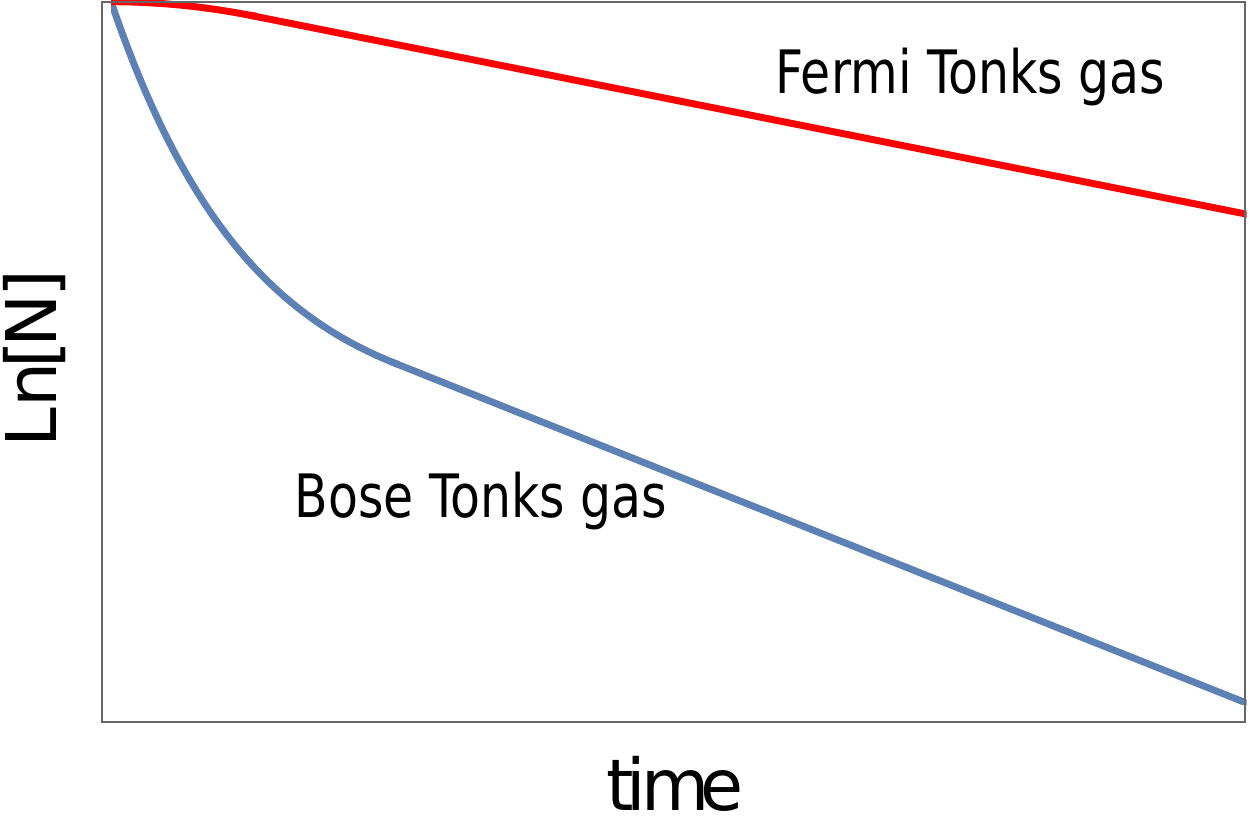}
\label{fig:schematic:b}
 }
 \hfill
 \subfloat[]{
 \includegraphics[width=7.8 cm]{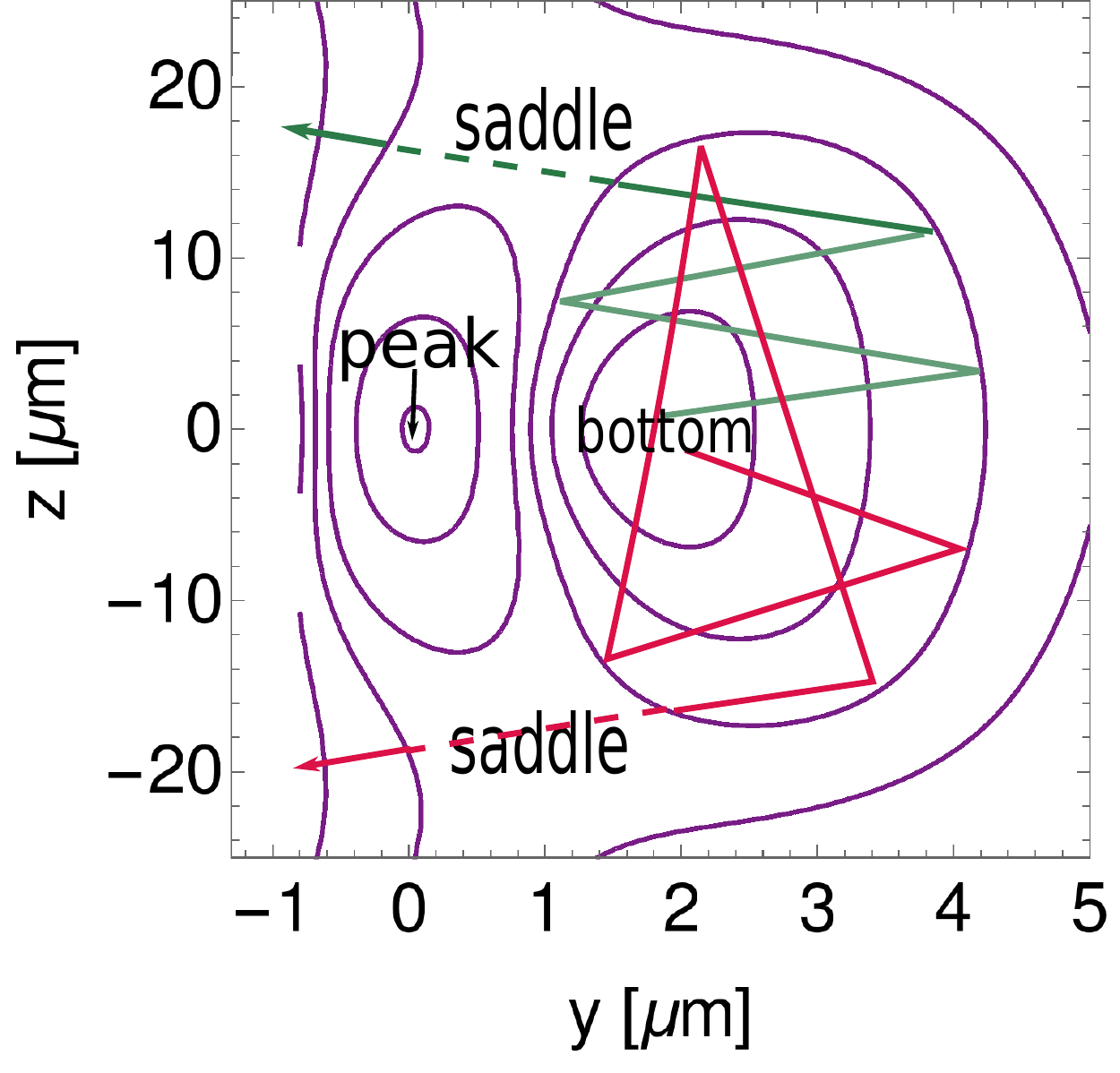}
\label{fig:schematic:c}
 }
 \caption{\textit{Novel Considerations Beyond Single-particle Tunneling.}  (a) Interatomic interaction reforms a square barrier, creating an effective potential. (b) Statistical properties of particles in Tonks-Girardeau gas, fermion (red) vs boson (blue), produce different short-time deviations in particle number from single-particle exponential decay.  (c) Trajectories of atoms in the trapping potential can become chaotic for many geometries including ours.  Despite these complications, effective 1D models can still be useful, as we demonstrate in this Article.}
\label{fig:schematic}
\end{figure}

There are several factors that can affect the behavior of quantum tunneling, which may push the tunneling process into distinct tunneling regimes. These factors include, but are not limited to, the following. First, the statistical properties of the trapped particles are a primary concern, e.g. whether exploring bosonic, fermionic, or anyonic tunneling~\cite{fermion2013,TGgas2006}, and whether the system can be treated as few-body or macroscopic. Second, interaction between particles -- varying from zero to non-zero~\cite{SFBCSBEC2012,natphy2013,PhysRevA.68.053614,PhysRevA.59.620,PhysRevB.78.041302}, weak to strong~\cite{mqt_bec2001, Carrboson2005,fragment1D2008} and attractive to repulsive~\cite{mqt_attactive_boson1998,Diego2016} -- may suppress or enhance tunneling, or even reform the barrier~\cite{dynamict2013}. Third, the shape of the potential well and assorted lattice and other geometries induce and alter quantum phases, such as in spin chains and disordered systems~\cite{qphase2007,rmpqphase2008}. Fourth, the dimension of the system can give rise to extra degrees of freedom which can for instance introduce chaos in the tunneling dynamics~\cite{chaos2007,fluctuation2002}.

These factors, together with the notion of macroscopicity, delineate the extensive territory of MQT into distinct tunneling behaviors. The term macroscopic may connect to an intuitive concept of ``large'' objects.  In fact the ``macroscopic'' in MQT has abundant aspects beyond one's first impression.
In the following, we lay out some of these possibilities, as a motivation for our own particular investigation, and to inspire future experiments and theory on the plethora of MQT phenomena, mostly still unexplored.

\subsection{Macroscopicity, Statistical Properties, and Quasiparticles}\label{section:statistic}

Macroscopicity, which is a prominent characteristic of MQT, features at least four facets. First, macroscopicity separates phenomena in a semiclassical phase space, but these phenomena are not necessarily describable by classical mechanics.  However, sometimes an extension of the semiclassical limit is apt.  For example, the phase-space analogy for a driven double-well BEC in a quantum two-mode approximation suggests that it can be mapped to a macroscopic superposition state of two pendulum rotor states~\cite{macrophase}. Second, macroscopicity can refer to a large quantity of particles, can activate many degrees of freedom, and can lead to emergent behavior. One example is ultracold bosons on a ring, where from the weakly interacting semiclassical limit in terms of dark solitons to the strongly interacting limit in terms of Yrast states, phase coherence can break down and phase slip enables continuous winding and unwinding of the system~\cite{soliton2008,soliton2009,soliton2010}. Third, macroscopicity can be associated with the notion of massive objects, where quantum mechanics meets gravity and may lead toward quantum gravity. One theoretical avenue is the Schr\"odinger-Newton equation, the Schr\"odinger equation with an additional gravitational potential term~\cite{SNeq,Penrose1996,SNeq1998}. Fourth, macroscopicity may exemplify complexity, which is garnering wider attention. For instance, recent studies of complex networks relying on quantum mutual information examine critical points for transverse Ising and Bose-Hubbard models~\cite{David2016}.

Tunneling bears a close relation to transport.  Loosely speaking, transport is above the barrier (often due to a ``push'') whereas tunneling is through the barrier.  Thus, one can consider tunneling of mass, charge, spin, etc.  Quantum tunneling is applicable far beyond the original concept of particle tunneling. For instance, we may speak of tunneling of magnetization in ferromagnetic films, or in a spin-1 BEC.
Studies of the transverse spin wave, with and without superflow, show different conditions for a certain anomalous tunneling behavior where tunneling occurs without reflection~\cite{spinwave2009, spinwave2011}.
We can also consider the tunneling between molecular states, such as ammonia and other pyramidal molecules~\cite{pwanderson,ammonia2014,pyramidal2002}, and the Josephson junction serves as the simplest possible model of such processes~\cite{zapata1998}.

The statistical properties of particles can be divided into fermions and bosons, and in lower dimensions include anyons; whether they contain only a single component or are a mixture of components; and whether the basic object exhibiting MQT is best described as emergent quasi-particles like vortices, solitons, or skyrmions.
Most such possibilities are accessible in cold quantum gases. The studies of few-particle system have already paved the way to some of these distinctions, especially between bosons and fermions. Bose and Fermi Tonks-Girardeau gases show different deviations from exponential decay at short times as a direct consequence of different ground state energies~\cite{TGgas2006}, shown in Fig.~\subref*{fig:schematic:b}. In few-fermion tunneling, pairing needs to be considered, a distinction from the boson case~\cite{fermion2013}. The influence of the inter-atomic interaction in these systems is also investigated, details in Section~\ref{section:interaction}. Non-exponential decay of ultracold single-atom tunneling is expected to occur in certain parameter regimes~\cite{Roberto2016}.

The tunneling of a BEC mixture or multi-component BEC also broadens the possibilities for MQT. A mixture can be divided into several cases: several different atomic species, each Bose-condensed; different internal states of the same atomic species, such as different hyperfine states of $^{87}$Rb; or use of different external states of a trap, in for instance MQT of quantum vortices~\cite{carr2011e}. For example, a study of MQT describes the mixing of two weakly-linked superfluids of interacting fermions, making it possible to obtain atomic Josephson junction equations describing the system as a whole~\cite{mixture2008}.  In general the MQT of domain walls, skyrmions etc. in such mixtures may carry different information and different decoherence properties from a bulk BEC.

Understanding the nature of the particles involved is a vital step in understanding any tunneling regime. Calculations can be remarkably simplified, for instance, if a quasi-particle representation is valid, or if a many-body system can be analyzed on a single-particle basis or a single-particle-like basis. Sometimes, a quasiparticle also induces new phenomena. For instance, the tunneling of nonequilibrium quasiparticles through a Josephson junction brings decoherence and produces energy decay in superconducting qubits and resonators~\cite{quasiparticle,quasiparticle2}. However, strongly-interacting systems may not exhibit quasiparticle representations, as we will discuss in the next section.

\subsection{The Role of Interactions} \label{section:interaction}

Inter-atomic interaction poses crucial considerations for quantum tunneling. A well-known example is Josephson dynamics~\cite{natphy2013,PhysRevA.68.053614,PhysRevA.59.620,PhysRevB.78.041302}, which occurs in systems of weakly coupled macroscopic quantum states. For weak interaction strengths, tunneling dominated effects such as the AC and DC Josephson effects~\cite{ACjosephson2005,DCjosephson2007} and coherent temporal oscillations~\cite{JJoscillation2002} occur, where condensates tunnel between two wells. When the interaction exceeds a critical value, the populations become self-trapped~\cite{PhysRevA.59.620}, where tunneling is suppressed and condensates are mainly located in one potential well. Similar effects also happen in exciton and polariton condensates~\cite{PhysRevB.78.041302}.  However, by biasing a more strongly-interacting system it is still possible to reduce tunneling times by tens to hundreds of orders of magnitude~\cite{carr2007b,carr2010i}.

Repulsive and attractive inter-atomic interactions can lead to different tunneling dynamics. Again, few-body tunneling presents several precursors to these effects in many-body systems. Fermionization of two distinguishable fermions occurs during the tunneling dynamics of a repulsively interacting system~\cite{Zurn2012,Massimo2012,Blume2015}. Sequential single-particle tunneling is found in both repulsive and attractive systems~\cite{Lundmark2015}, while pairing phenomena can be investigated in the strongly attractive interaction region~\cite{fermion2013,Lundmark2015,Blume2015}. As inter-particle interaction varies from strongly attractive to strongly repulsive, tunneling rates diverge within a wide range of orders of magnitude~\cite{Lundmark2015,Blume2015}.

The strength of the interaction in a system determines the appropriate tunneling theory. For instance, bosons with weak interactions can often be described by mean-field theory, where the many-body wavefunction is dominated by a semi-classical field, i.e., a complex scalar field~\cite{mqt_attactive_boson1998, mqt_bec2001, Carrboson2005}. Even weak interactions produce an effective barrier, which deforms as the wavefunction escapes, as depicted by the blue curve in Fig.~\subref*{fig:schematic:a}, where repulsive atomic interactions change the barrier the wavefunction encounters, resulting in a non-exponential decay.

Increased interaction strength can lead to correlations, fluctuations, and entanglement~\cite{interaction}, which render mean-field theory ineffective. We include here a non-exhaustive list of resulting effects on MQT.  First, the BEC may be fragmented or depleted. When the energy of the system exceeds a threshold as a consequence of strong interactions, fragmentation -- macroscopic occupation of more than one mode of the single-particle density matrix -- is induced~\cite{fragment1D2008}. The tunneling of an initially parabolically-trapped ultracold Bose gas in a coherent state into open space develops fragmentation after some propagation time~\cite{fragment_openspace2012,Diego2016}; the fragmented components are not said to be phase-coherent, as the notion of phase is tied to a single macroscopically occupied mode. Thus semiclassical wave theory becomes less and less applicable, and tunneling does not proceed in the same manner as the original concept.  Nevertheless, particles with an energy below that of a potential barrier tunnel, so the term, even in this non-semiclassical context, still applies.  Apart from fragmentation, condensates can also deplete, where there is non-macroscopic occupation of many modes~\cite{Diego2016}.  Second, interactions can enhance or decrease tunneling rates. The tunneling rate of a quasibound many-body state is sped up (slowed down) by repulsive (attractive) interactions~\cite{Diego2016}. Third, fluctuations affect the tunneling rate. Quantum fluctuations of the Josephson-Leggett mode in a Josephson junction drastically enhances the MQT escape rate~\cite{PhysRevB.83.060503,PhysRevB.89.224507}. Finally, dissipation may also alter the tunneling behavior. In the Josephson-Leggett mode, quantum dissipation suppresses  MQT~\cite{PhysRevB.89.224507}.

For strong interactions, 1D bosons can show similar transport properties as noninteracting fermions, due to the boson-fermion dual representation in 1D~\cite{PhysRevA.84.013608}. A famous example is a Tonks-Girardeau gas, a strongly interacting system, where bosonic atoms exhibit fermion-like behavior~\cite{TGMG1960}.  A non-exponential decay regime arises in a bosonic Tonks-Girardeau gas for short times, which exhibits few-body decay features. A fermionic Tonks-Girardeau gas, on the other hand, shows bosonization and deviations from exponential decay at long times~\cite{TGgas2006}.  As an extreme example of strong interactions, consider the unitary Fermi gas, and its holographic dual in a weakly curved gravitational representation~\cite{carr2012c}.  The MQT of a unitary Fermi gas remains an exciting and untested prediction of dynamical holographic approaches~\cite{Adams2014}.

\subsection{Trapping Potential}

The potential dictates the tunneling environment, and plays an essential role in characteristic tunneling times. An important case across many fields of physics, chemistry, etc. is tunneling from one or a few single-well discrete modes to continuous modes in free space, i.e., the quasi-bound or escape problem.  This is exactly the system studied in this Article, where in our case tunneling escape from a single well is assisted by inter-atomic interactions~\cite{shreyas2016}. For a  harmonic trap, power law behavior in MQT of BECs is observed near the critical point of collapse, while for an anharmonic trap, there is no power law behavior~\cite{Haldar2013}.

Double well tunneling stands in strong contrast to the quantum escape problem, since for instance a small bias in the system can lead to an exponential suppression of tunneling. Both wells contain discrete modes for atoms, thereby the tunneling process describes motion from discrete modes in one well to discrete modes in the second well. Josephson effects~\cite{natphy2013,PhysRevA.68.053614,PhysRevA.59.620,PhysRevB.78.041302,ACjosephson2005}, including the well-known AC and DC Josephson effects~\cite{ACjosephson2005, DCjosephson2007}, which arise in double-well-type systems, epitomize this form of MQT. Related contexts in this direction range from adiabatic transport of BECs ~\cite{Nesterenko2009} to interaction supported transport of BECs~\cite{Nesterenko2014}, and polarized fermion tunneling in 3D~\cite{fermion3D2013}. Under common assumptions, only a few modes in each well are required for consistent analysis with experiments.

In addition to these examples, other potentials to consider include periodic potentials, like lattices. In a static periodic lattice potential, the expansion of matter waves is quadratic for short times~\cite{lattice2010}. Other than potential shapes, barrier materials also modify the tunneling process. Comparing two different barrier materials, pillared silicon and aerogel, quantum reflection is suppressed differently by mean-field interactions at low velocity~\cite{pmaterial}. In addition to the bare potentials above, there are also dressed potentials, such as radio frequency dressed potentials~\cite{dressedp}. Studies show that on-site interactions in neutral atoms are dramatically enhanced in radio frequency dressed traps, which can affect the tunneling behavior~\cite{rfdressed}.

Beyond static potentials, dynamic potentials, such as those directly influenced by a time-dependent force, provide a unique milieu for tunneling dynamics. A ring trap with a driven space-time reflection symmetry breaking (PT symmetry) potential can induce chaos in the system~\cite{drivenPT2013,carr2016d}. Shaking a lattice can reduce the tunneling rate, and even completely suppress it for certain values of the shaking or driving parameter~\cite{shaking2007,driven2007,lattice2008,lattice2010}; the same phenomenon also occurs in a driven double well system~\cite{driven1991}. In a single potential well with a periodically modulated amplitude, dynamical tunneling is suppressed by nonlinear interaction, in comparison to the non-interacting case~\cite{PhysRevLett.109.080401}. A less obvious dynamic barrier can form indirectly by other means. For instance, interactions between particles can deform the barrier, rendering a previously static barrier dynamic, as shown in Fig.~\subref*{fig:schematic:a}.

\subsection{Dimension of the System}
Tunneling usually occurs at the weakest points of a trap, where the escape rate is the largest and most favorable. For instance, in this Article, tunneling proceeds through the weakest points of the potential well, the two saddle points as depicted schematically in Fig.~\subref*{fig:Torontoexp:b}. This is similar to a dam breaking; water begins to spill out from the structurally weakest points. In this view, tunneling can be recast as a 1D or quasi-1D problem. But, higher dimensions can elicit new phenomena. One such phenomenon is chaos, which can significantly complicate the dynamics~\cite{chaos2007}. The cluttered paths in Fig.~\subref*{fig:schematic:c} sketch possible chaos in the experimental trap, and dashed lines at the saddle points indicate MQT. Chaos-assisted tunneling sometimes gives rise to tunneling oscillation~\cite{fluctuation2002}, and produces irregular fluctuations in the tunneling rate~\cite{fluctuation2001,chaos1994}. Another consequence, due to the presence of extra degrees of freedom, is incoherent oscillations of a large number of polarized fermions in a 3D double well~\cite{fermion3D2013}.

\section{A Macroscopic Quantum Tunneling Experiment}\label{section:exp}

We have made the first step in reporting a non-exponential decay in a single-well tunneling experiment due to inter-atomic interactions~\cite{shreyas2016}.  In this section, we offer a description of the experiment, present previously unpublished findings from both the experiment and supporting 3D mean-field simulations showing reasonable agreement with the data, and finally exhibit a case study of the experiment using an effective 1D modified JWKB model derived in detail in Section~\ref{section:WKB}.

\subsection{Experimental Design}\label{section:expDesign}

The experiment studied tunneling of a $^{87}$Rb BEC prepared in a quasi-bound state with repulsive inter-atomic interaction. The trapping potential had harmonic confinements in the $x$ and $z$ directions with trapping frequencies $\omega_x$ and $\omega_z$. Due to gravity and a magnetic field gradient, there was a tilt in the $y$ direction (vertical direction) with a constant acceleration $a$. A slice through the $y$--$z$ plane of the potential well is shown in Fig.~\subref*{fig:Torontoexp:b}, with complete trapping potential as follows:
\begin{eqnarray}
V(x,y,z)&=&\frac{1}{2}m\omega_x^2x^2+\frac{1}{2}m\omega_z^2z^2-may+V_\mathrm{b}(x,y,z)\nonumber\\
V_\mathrm{b}(x,y,z)&=&V_0\exp\left(-2y^2/\omega(z)^2\right)\nonumber\\
\omega(z)&=&\omega_0\left(1+\left(z/z_\mathrm{R}\right)^2\right)^\frac{1}{2}
\label{eq:trap}
\end{eqnarray}
Here ${V_0}$ is the peak height of the barrier, $z_{\rm R}=8\,\mu$m is the Rayleigh range, $\omega_0$ is the barrier waist, $\omega_0=1.3(0.1)\,\mathrm{\mu m}$, and $\omega(z)$ is the Gaussian beam waist.

The experiment uses two trapping configurations: weak and tight. The parameters of the trap in the weak configuration are $\omega_x=2\pi\times32.7 (0.24) \rm{\,Hz}$, $\omega_z=\omega_x/2$, and $a=2.08 (0.04)\mathrm{\,m/s^2}$, while for the tight configuration, $\omega_x=2\pi\times86.6 (0.6) \rm{\,Hz}$, $\omega_z=\omega_x/2$, and $a=8.40 (0.06)\mathrm{\,m/s^2}$. As a result of a tighter confinement in the latter, the initial total number of atoms trapped in the potential is around one fifth of that in the weak configuration. This decrease of particle number affects the validity of the Thomas-Fermi approximation in our previous mean-field calculation~\cite{shreyas2016}, and requires an offset of the kinetic energy term. The details of this correction is described in Section~\ref{section:2nd}. Although the first report of our experiment focused on the tight trapping configurations~\cite{shreyas2016}, our theory starts with the weak configuration and then covers the tight configuration afterward, expanding the reach of our study to different regimes.

Figure~\subref*{fig:Torontoexp:a} depicts the time sequence of the experimental procedures. First, a cloud of $^{87}$Rb atoms in the $\mid F=2,m_{F}=2\rangle$ ground state is loaded into the trapping potential from a hybrid trap. Next, evaporative cooling techniques lower the temperature of the atoms until they form a BEC. Then, the barrier height is ramped down non-adiabatically in 20\,ms (5\,ms) for the weak (tight) configuration. After ramping procedure, the condensate is held in the trap for a variable time from 0.1\,ms to 1.2\,s. Finally, the trapping potentials are abruptly turned off and the cloud is imaged after time-of-flight (TOF) expansion.

\begin{table}
\begin{tabular}{| c | c | c | c |}
\hline
Peak [nK] & Saddle [nK] & $x_{0}$ [$\mathrm{\mu m}$] & $N_{0}/10^3$\\
\hline
$146 \pm 10$ & $ 46 \pm 3 $ & $16 \pm 1$ &$423 \pm 21$\\
$170 \pm 11$ & $ 52 \pm 10 $& $17 \pm 2$ &$548 \pm 37$\\
$190 \pm 13$ & $ 58 \pm 4 $ & $18 \pm 1$ &$559 \pm 33$\\
$221 \pm 16$ & $ 69 \pm 5 $ & $20 \pm 1$ &$630 \pm 26$\\
$260 \pm 18$ & $ 81 \pm 6 $ & $21 \pm 1$ &$583 \pm 32$\\
$299 \pm 21$ & $ 92 \pm 6 $ & $23 \pm 1$ &$663 \pm 54$\\
\hline
\end{tabular}
\caption{\label{table:exptable}\textit{Experimental and Estimated Parameters.} From the weak configuration, with peak height, saddle height, horizontal distance from local potential minimum to saddle point ($x_{0}$) and initial total particle numbers trapped behind the barrier ($N_{0}$).}
\end{table}

Table~\ref{table:exptable} shows major parameters in the weak configuration. The relative saddle heights are about one-third of their corresponding peak heights, where we note that MQT occurs through the two saddles or weak points of the potentials, not the peak. All the plots in this Article reference peak heights. In the trap, $x_{0}$ is the horizontal distance from the bottom (local minimum point) to one saddle point. We will see in Section~\ref{section:WKB} that our theoretical fit parameters agree well with this parameter. $N_{0}$ is the total particle number in the trap at the beginning of experimental observation of the dynamics, after the non-adiabatic ramp down of the barrier.  Figure~\subref*{fig:Torontoexp:c} shows a typical raw data set from the experiment.  We find clear non-exponential decay in the number of atoms.  Further presentation and analysis of the data is deferred to the following two sections.

\subsection{Experimental Data and 3D Mean-Field Model}\label{section:3Dsimu}

\begin{figure}
 \subfloat[]{
 \includegraphics[width=8.3 cm]{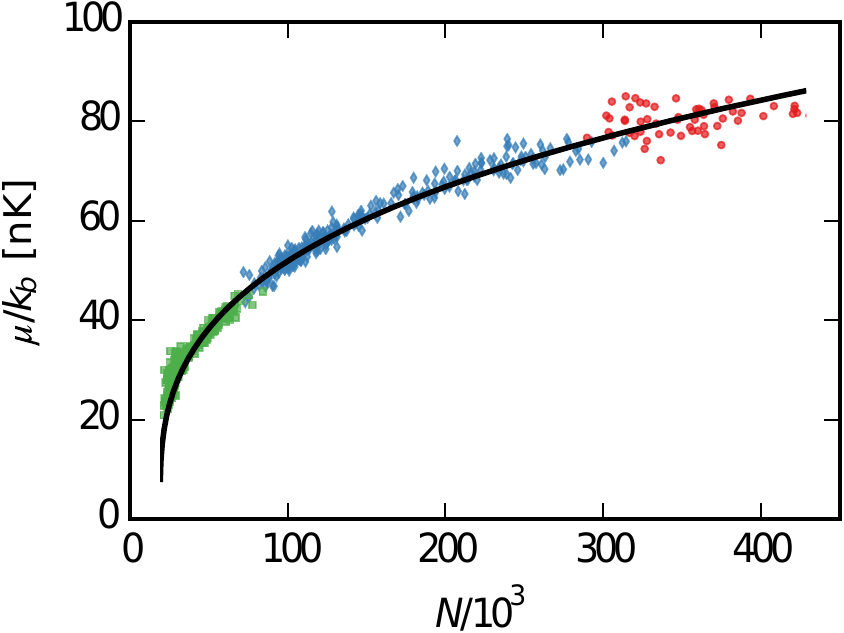}
 \label{fig:mu3D:a}
 }
 \hfill \\[-1ex]
 \subfloat[]{
 \includegraphics[width=8.3 cm]{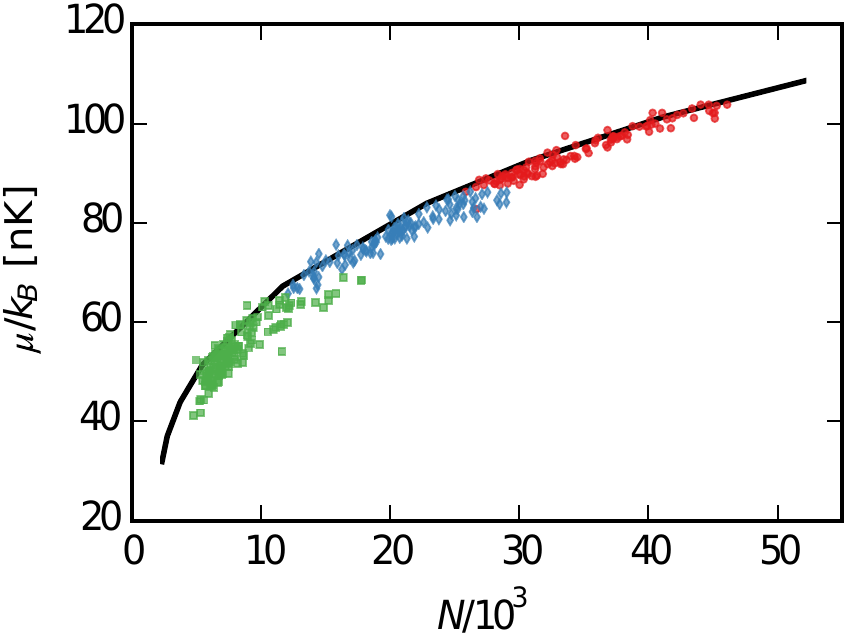}
 \label{fig:mu3D:b}
 }
\captionsetup{justification=justified,singlelinecheck=false}
\caption{\textit{Experimental and Numerical 3D Gross-Pitaevskii Chemical Potential.} Chemical potential $\mu$ as a function of the total number of atoms $N$. (a) Weak configuration barrier heights: $V_{0}$ = 460(30)\,nK (red circles), 260(18)\,nK (blue diamonds), and 170(11)\,nk (green squares). (b) Tight configuration: $V_{0}$ = 330(35)\,nK (red circles), 290(30)\,nK (blue diamonds), and  240(25)\,nK (green squares). Mean-field simulations predict a chemical potential (black line) in agreement with experimental data, using barrier heights (a) 350\,nK and (b) 300\,nK.}
\label{fig:mu3D}

\end{figure}

\begin{figure}
 \subfloat[]{
 \includegraphics[width=8.3 cm]{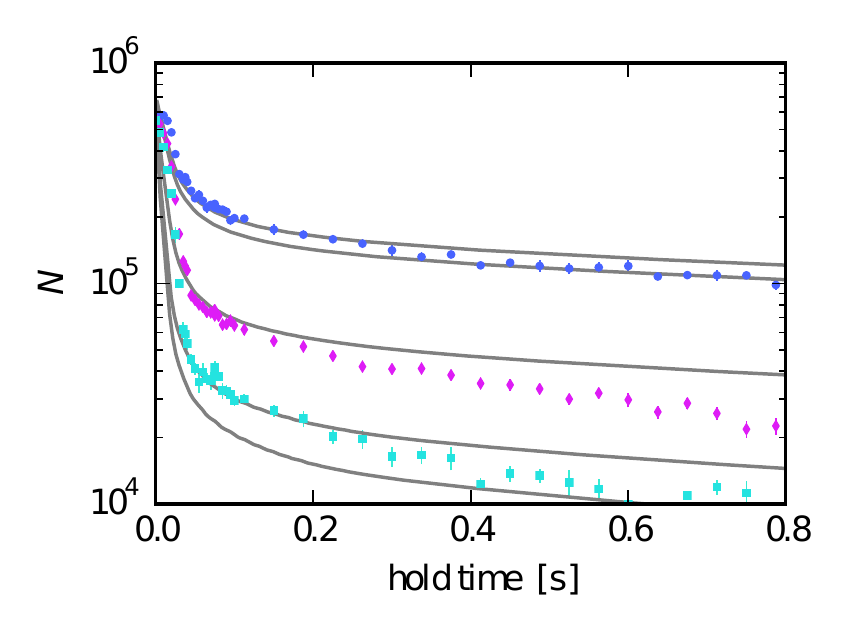}
 \label{fig:Nt3D:a}
 }
 \hfill \\[-3ex]
 \subfloat[]{
 \includegraphics[width=8.3 cm]{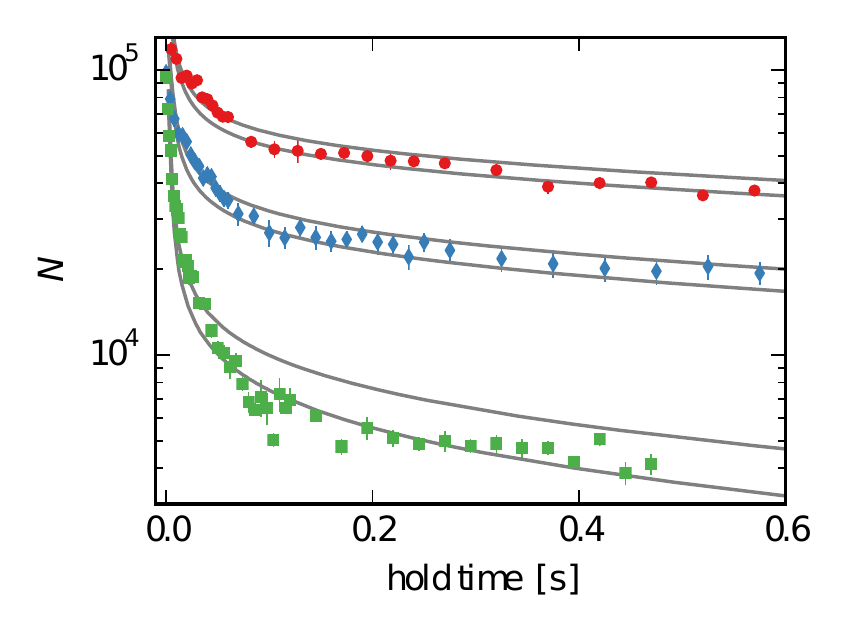}
 \label{fig:Nt3D:b}
 }
  \hfill \\[-3ex]
 \subfloat[]{
 \includegraphics[width=8.0 cm]{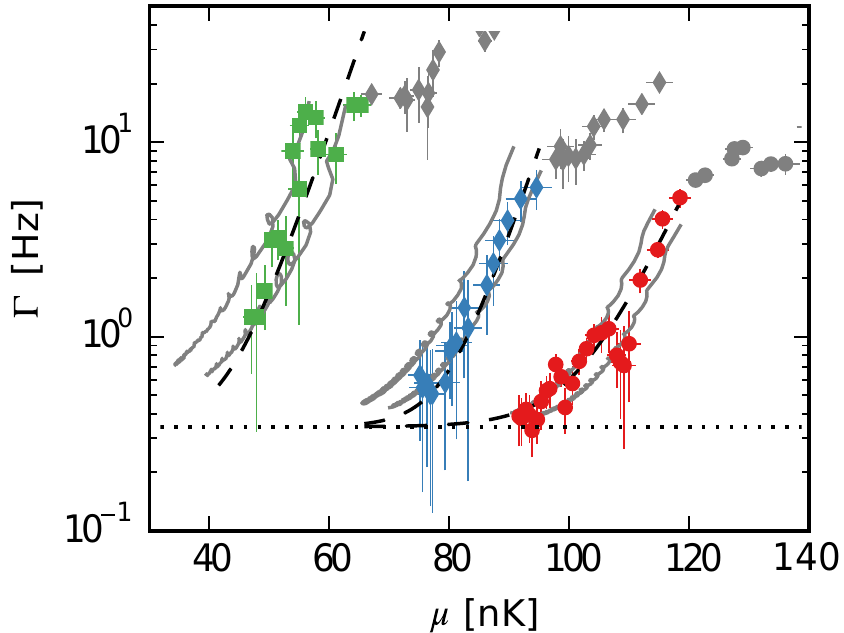}
 \label{fig:Nt3D:c}
 }
\caption{\textit{Numerical 3D Gross-Pitaevskii and Experimental Data.} Semi-log plot for decay in number of trapped atoms for (a) weak and (b) tight trapping configuration. Data from the classical spilling regime around the first 40 ms (20 ms) in weak (tight) configuration is also included. Gray lines are 3D GPE simulations. Beyond (a) 0.8\,s and (b) 0.6\,s, the dynamics become loss dominated. (c) 3D simulations reproduce gross features of decay rate as a function of chemical potential, where gray data points are classical spilling transients, and dashed black lines are fitting results. All data points are from the experiment.}
\label{fig:Nt3D}
\end{figure}

Data for the chemical potential as a function of atom number and time dependence of the number of atoms trapped in the quasi-bound state are shown in Fig.~\ref{fig:mu3D} and~\ref{fig:Nt3D} for both weak and strong trapping configurations.  Mean-field simulations are then performed for comparison, using the split-step operator method for the 3D Gross-Pitaevskii Equation (GPE); imaginary time evolution is used to find the ground state.  Measured trap parameters and initial atom numbers are used as input parameters to the simulations without any free parameters. To mimic the experimental procedure in Fig.~\subref*{fig:Torontoexp:a}, the barrier height is linearly ramped down to a final value. Absorbing boundary conditions are introduced to avoid reflections of escaped atoms from the edge. Finally, the chemical potential extracted from the experiment uses a Thomas-Fermi approximation~\cite{shreyas2016}.  This leads to good agreement in the weak trapping case in Fig.~\subref*{fig:mu3D:a} but a systematic error in Fig.~\subref*{fig:mu3D:b} where the Thomas-Fermi approximation is less applicable due to a larger kinetic or zero-point energy.  In contrast, in Fig.~\ref{fig:Nt3D} the mean-field time sequence is qualitatively close but is up to a factor of 2 off in both trapping configurations.
These studies suggest a mean-field model can reproduce the main features of this experiment, but may bear some correcting either due to characterization of the trapping potential or due to other beyond mean-field effects, as we will explore in Sec.~\ref{section:TEBD}.

\begin{figure*}
\begin{center}
\includegraphics[scale=0.25]{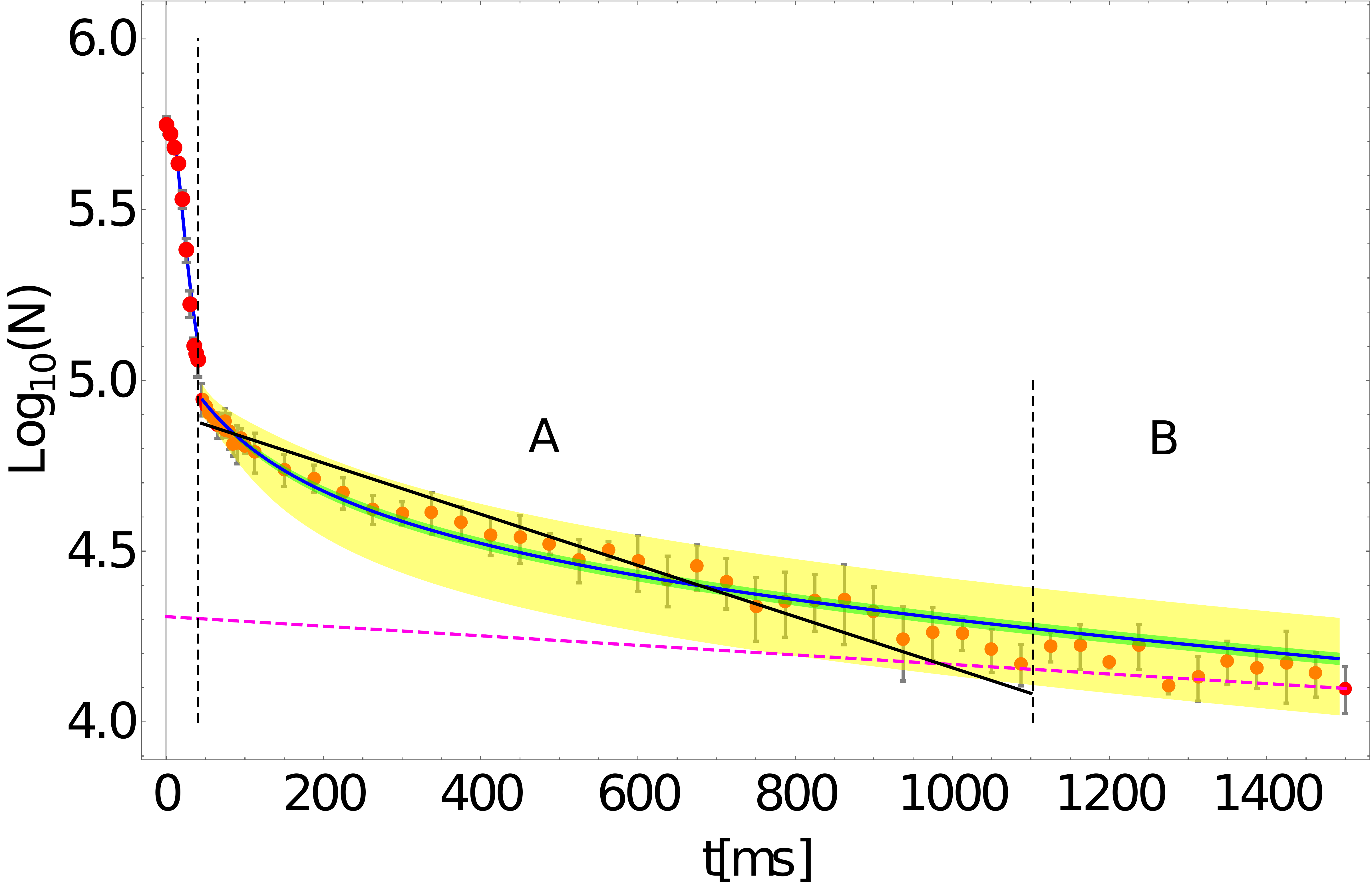}
\end{center}
\caption{\textit{Case Study: Theoretical Model Fit to Macroscopic Quantum Tunneling Data.} Experimental number of trapped atoms as a function of time (red circles: mean values with 1$\sigma$ error bars) with theoretical fit (blue curve), exponential fit through tunneling dominated regime (solid black line), and experimental background loss (dashed pink nearly-horizontal line) We divide the decay curve into three sub-regions (indicated by dashed vertical lines): initial transient classical spilling, (A) mean-field assisted quantum tunneling region with non-exponential decay, and (B) background loss dominated region. The green envelope indicates uncertainty in fitting parameters from modified JWKB.  The yellow envelope indicates combined uncertainty due to: uncertainty in experimental parameters, uncertainty in data, and uncertainty in fit parameters.}
\label{fig:fig186}
\end{figure*}

Both experimental data and mean-field predictions clearly show non-exponential decay in both trapping configurations.  Figure~\subref*{fig:Nt3D:a} covers the weak configuration with  barrier heights $V_{0}$ = 260(18)\,nK (blue circles), $V_{0}$ = 190(13)\,nK (purple diamonds) and $V_{0}$ = 170(11)\,nK (teal squares), and Fig.~\subref*{fig:Nt3D:b} the tight configuration with barrier heights $V_{0}$ = 330(35)\,nK (red circles), $V_{0}$ = 290(30)\,nK (blue diamonds) and $V_{0}$ = 240(25)\,nK (green squares). From bottom to top, 3D GPE simulations are shown by the solid gray lines with barrier heights for Fig.~\subref*{fig:Nt3D:a}: 120, 130, 140, 210, and 220\,nK; and in Fig.~\subref*{fig:Nt3D:b}: 230, 240, 290, 300, 340 and 350\,nK. The barrier heights used in the simulations were chosen to match the results from the experiment. In the case of the tight configuration (Fig.~\subref*{fig:Nt3D:b}), we see that the barrier heights for which the simulation closely matches the data are consistent with the experimentally measured barrier height. However, in the case of the weak configuration (Fig.~\subref*{fig:Nt3D:a}), there is a discrepancy. This could be possibly be due to aberrations developed in the barrier beam off-axis, which results in an incorrect estimation of the barrier height. Quantum tunneling starts at around 40 ms (20 ms) in weak (tight) configuration. The non-exponential decay feature of this quantum tunneling region (shown in Fig.~\subref*{fig:Nt3D:a} and Fig.~\subref*{fig:Nt3D:b}) will be emphasized in Fig.~\ref{fig:fig186}.

We further consider the relation between decay rate and chemical potential in Fig.~\subref*{fig:Nt3D:c} for the tight trapping configuration, with barrier heights of $V_{0}$ = 240(25)\,nK (green squares), $V_{0}$ = 290(30)\,nK (blue diamonds), and $V_{0}$ = 330(35)\,nK (red circles).  Their corresponding 3D mean-field simulations are shown for barrier heights of 230 and 240\,nK, 290 and 300\,nK,  340 and 350\,nK.  The 3D mean-field simulations fit the tight configuration well, which did not exhibit strong classical spilling in the initial stages of the dynamics, but not the weak trapping configuration, which had well-identified distinct classical tunneling and quantum tunneling regimes, and is discussed further in Sec.~\ref{section:WKB}. We observe that the decay rate can be fitted with a simple exponential function of the chemical potential of form
\begin{equation}
\Gamma=\Gamma_{\rm bg}+\exp (\alpha+\beta \mu)\label{eq:decayrateexp}.
\end{equation}
Here $\Gamma_{\rm bg}$ is the background loss rate, and $\Gamma_{\rm bg}=0.31(0.02)\,\mathrm{Hz}$ in the experiment. These fits are included in Fig.~\subref*{fig:Nt3D:c} as dashed black lines.

\subsection{Effective 1D JWKB Description: An Experimental Case Study}
\label{section:case_study}

In the weak configuration, there is spilling dominated dynamics for approximately the first 40\,ms, quantum tunneling from about 40\,ms to about 1\,s, and background loss dominates thereafter. While in the tight configuration, there is spilling dominance until around $0\sim20\,\mathrm{ms}$, quantum tunneling from 20\,ms to 1.1\,s, and experimental background loss thereafter. During the escape, trapped atoms tunnel from a quasi-bound state into the continuum via the weakest points, the saddle points, as the rate is largest and favorable. Figure~\subref*{fig:Torontoexp:b} schematically shows the corresponding representative trajectories.  An intriguing question is whether or not, in this rather complicated potential possibly supporting chaotic semiclassical trajectories, one can capture the basic dynamics of MQT with the well-known JWKB model in an effective 1D picture.  To explore such an idea, as illustrated in Figure~\ref{fig:Torontoexp}, we choose the most probable direct path for MQT, from the local minimum behind the barrier to the saddle points.

Figure~\ref{fig:fig186} presents one case study with a barrier height of 190(13)\,nK   in the weak configuration and serves to convey the details of our approach. We divide the dynamical process into three main parts: transient spilling, mean-field assisted MQT, and background loss dominated dynamics, division indicated by the vertical dashed lines. In our analysis we discard the classical spilling transients.  MQT is defined as beginning when the chemical potential is equal to $V_{\rm s}$, the difference in potential energy between the saddle points and the local minimum of the potential. For the weak configuration, this is at about $t=20$ to 40\,ms, beginning region A.  In this region, the decay process slows down and shows a non-exponential decay feature; an exponential fit (solid black line) has $\chi^{2}=3.32$ as opposed to our model fit $\chi^{2}=1.21$. This non-exponential decay is caused by the mean-field effect, or atomic interactions, as confirmed by the effective JWKB model fit and the 3D GPE simulations.  As we show in Section~\ref{section:WKB}, this fit requires the interactions to match the data.  As the decay process progresses, the number of atoms and the chemical potential decreases, and so does the mean-field effect. Thus, we obtain an effectively dynamical barrier height, which decreases with time.  Combined, these factors produce a faster decay at the beginning of region A and a much slower decay by the end of the region. Finally, in region B, the decay process is dominated by the background loss. The blue curve is our theoretical fit and red points with error bars are the experimental data. The error region for our fit, from the uncertainty in our resulting fit parameters ($a$ and $w$ as discussed in Section.~\ref{section:WKB:subsection}), is shown in green. The error region shown in yellow is the combined error including uncertainty in (i) fit parameters, (ii) atom number (error bars for red points), and (iii) experimental parameters; errors added in quadrature.  The major contributions to the experimental error are the standard deviations of initial particle numbers ($\delta N_{0}$) at the start of the tunneling regime, and uncertainty in the peak barrier height ($\delta V_{\mathrm{s}}$), about 6\% (11\%) in the weak (tight) configuration. For early times, $t < 100 \mathrm{\mu s}$, both $\delta N_{0}$ and $\delta V_{\mathrm{s}}$ contributed about $O(10^{3} \sim 10^{4})$ to the total error envelope. After this time, the contribution from $\delta N_{0}$ decreases 1-2 orders of magnitude, while $\delta V_{\mathrm{s}}$ contribution remains about the same. All other error contributions are generally at least one order of magnitude smaller than $\delta V_{\mathrm{s}}$. We found similar results for the tight configuration, but the smaller particle numbers resulted in larger error envelopes relative to particle numbers; see Section.~\ref{section:2nd} for  details.  We also calculate the reduced chi squared of our model for all the data sets in both configurations, which confirms that our theoretical fits are well within experimental error.

In fact, this pattern in Fig.~\ref{fig:fig186} occurs in all the experimental data sets, in both weak and tight configurations, as we discuss in the following section.

\section{Effective 1D JWKB Model of Macroscopic Quantum Tunneling}\label{section:WKB}

In this section, we derive and explore in detail an effective 1D modified JWKB model, comparing quantitatively to experimental results.  The term ``modified'' refers to inclusion of mean-field effects not normally considered in based JWKB analysis, which turn out to be key to the non-exponential decay observed in the experiment.

\subsection{Modified JWKB Model}
\label{section:WKB:subsection}

\begin{figure}[t]
 \includegraphics[width=8.3 cm]{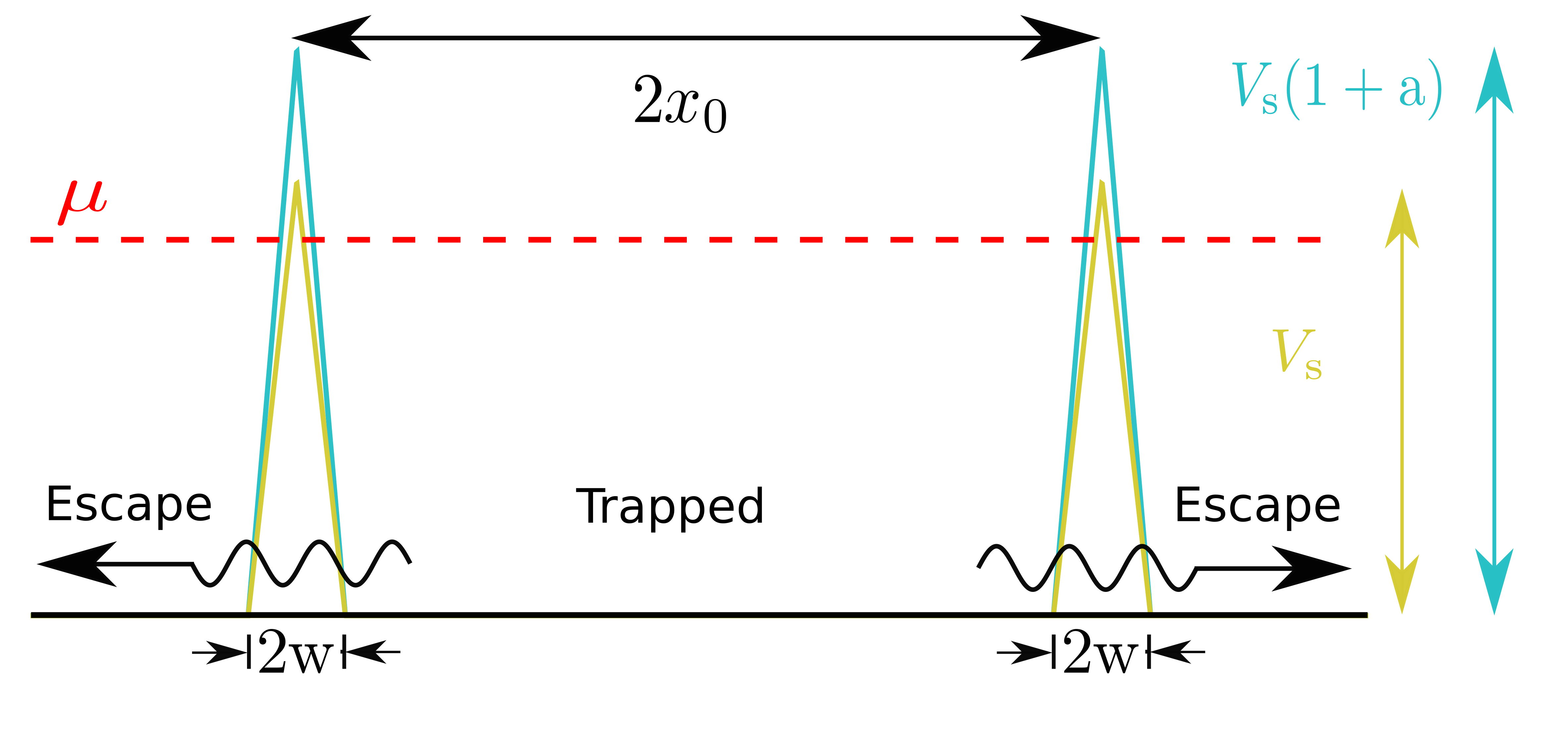}
\caption{\textit{JWKB Potential Schematic.} Tunneling in the full 3D experimental potential can be modeled with a much simpler 1D approach.  The minimal feature set required to fit the data is (1) a bare 1D triangle  potential of height $V_{\rm s}$ and width $2w$ (gold); (2) inclusion of the mean-field to create an effective potential of height $V_{\rm s}(1+a)$ with $a\propto N(t)$ (teal); and (3) use of the Thomas-Fermi approximation for the chemical potential $\mu\propto N^{1/3}$ (red dashed line).  Such a choice of model provides self-consistency between the replacement of $E$  with $\mu$ in the usual JWKB treatment, together with the bare $V(x)$ with the effective potential $V_\mathrm{eff}(x)=V(x)+g|\psi|^2$.}
\label{fig:tri_schematic}
\end{figure}

Our initial hypothesis was that we could capture non-exponential tunneling simply through barrier shape.  A triangle barrier is the simplest case with a barrier width which increases as the particles escape and the chemical potential decreases, leading to a slow-down in tunneling.  The triangle also has the advantage of being analytically tractable. We found that a square barrier does not reproduce the experimental data, while a Gaussian increases the analytical difficulty without improving the accuracy of the model.  We take the triangle height to be the saddle height $V_{\mathrm{s}}$ and triangle centers at $x=\pm x_0$, both as determined in the experiment, and single free parameter, the full width at half maximum (FWHM), $w$ (i.e., its slopes are $\pm V_{\mathrm{s}}/w$):
\begin{equation}
	V_{\mathrm{tri}}(x) =
	\left\{
	\begin{matrix}
		-\frac{V_{\mathrm{s}}}{w}|x\pm x_{0}|+V_{\mathrm{s}},  & |x\pm x_{0}| < w\\

		0,& \text{otherwise}\\
	\end{matrix}
	\right. \,\\
\label{eq:triangle}
\end{equation}
as shown in Fig.~\ref{fig:tri_schematic}.
Using Eq.~(\ref{eq:triangle}), the JWKB tunneling probability is given by
\begin{eqnarray}
P_{\mathrm{tri}}&=&\exp\left(-\frac{8}{3\hbar}w\sqrt{2mV_{\mathrm{s}}} \left(1-\frac{E}{V_{\mathrm{s}}}\right)^{3/2}\right),
\label{eq:WKBp}
\end{eqnarray}
the semi-classical oscillation time in the potential by
\begin{eqnarray}
\tau_{\mathrm{tri}}&=&\sqrt{2m}\frac{(x_0-w)}{\sqrt{E}}+2\sqrt{2m}\frac{w \sqrt{E}}{V_{\mathrm{s}}},
\label{eq:WKBtau}
\end{eqnarray}
and the tunneling rate through the saddle by
\begin{eqnarray}
\Gamma_{\mathrm{tri}}=\frac{P_{\mathrm{tri}}}{\tau_{\mathrm{tri}}}. \label{eq:Gamma_tri}
\end{eqnarray}
The experimentally measured background loss rate is $\Gamma_{\mathrm{bg}}\approx 0.31(0.02)\,\mathrm{Hz}$.
The number of atoms remaining behind the barrier as a function of time is given by a decay equation
\begin{eqnarray}
\dot{N}(t)=-(\Gamma_{\mathrm{tri}}+\Gamma_{\mathrm{bg}})N(t) = -\Gamma N(t), \label{eq:decay_eq}
\end{eqnarray}
We further assume our system can be described by the Thomas-Fermi picture, which justifies using the analytical approximation for the chemical potential~\cite{shreyas2016}, from which we obtain
\begin{equation}
\mu = b N^{1/3}, \label{eq:chemical_arppox}
\end{equation}
with constant of proportionality $b = 1.15(8)$\,nK ($3.03(5)$\,nK) as measured for the weak (tight) configuration.  Note that any dependence on barrier height is negligible within experimental and model fitting parameter error.

The next logical step to adapt JWKB to this context is to replace $E$ in Eq.~(\ref{eq:WKBp}) and Eq.~(\ref{eq:WKBtau}) with the appropriate single-particle energy in the presence of the effective potential and interactions from Eq.~(\ref{eq:chemical_arppox}).
However, this approach alone fails to accurately reproduce the experimental results over the entire experimental time-window: tunneling stops too soon and the dynamics of the final half of the tunneling process are not accounted for. Instead, along with replacing $E$ with $\mu$ in Eq.~(\ref{eq:chemical_arppox}), the explicit inclusion of mean-field effects is also required to fit the data, which makes the tunneling rate $\Gamma$ in Eq.~(\ref{eq:decay_eq}) time dependent, similar to results from the complex scaling method~\cite{Schlagheck2007}.
In order to take into account the time-dependent mean-field effects, we add time dependence to the height of our triangle potential, capturing a mean-field effective potential from the GPE of form $V_\mathrm{eff}(x)=V(x) + g |\psi(x,t)|^2$, as is also necessary to be self-consistent with the Thomas-Fermi approximation underlying Eq.~(\ref{eq:chemical_arppox}); see also~\cite{Moiseyev2004,Carr2005}. The simplest model capturing such effects is a potential of form
\begin{align}
\begin{split}
	&V_{\mathrm{eff}}(x,t) =
	\left\{
	\begin{matrix}
		-\frac{V_\mathrm{mf}}{w}|x\pm x_{0}|+V_\mathrm{mf},& |x\pm x_{0}| < w,\\
		0, & \text{otherwise},\\
	\end{matrix}
	\right .\\
	&V_\mathrm{mf} = V_{\rm s} \bigl[1+ a \frac{N(t)}{N_0} \bigr].\\
\end{split} \label{eq:new_veff}
\end{align}
Here, $N_0$ is the initial particle number in the trap when quantum tunneling regime begins ($t_{\mathrm{QT}}$), $a>0$ is a unitless mean-field fit parameter, and $V_\mathrm{mf}$ is the time-dependent potential pre-factor that captures the mean-field. The effect of $a$ in $V_\mathrm{mf}$ is to introduce a dynamic effective barrier which decreases to the bare potential as the atoms escape. A schematic of the bare triangle barrier Eq.~(\ref{eq:triangle}) and effective mean-field triangle Eq.~(\ref{eq:new_veff}), at the beginning of the tunneling regime $V_\mathrm{eff}(x,t_{\mathrm{QT}})$, are shown in Fig.~\ref{fig:tri_schematic}.  Note, mean-field interaction would necessarily alter the ground-state wavefunction of the system; the JWKB model used here is a minimal model which is able to reproduce the experimental findings with only two free parameters, $w$ and $a$, and may hint at many-body dynamics as will be discussed in Section~\ref{section:TEBD}. Finally, we make the appropriate replacements of Eqs.~(\ref{eq:chemical_arppox}) and ~(\ref{eq:new_veff}) into Eqs.~(\ref{eq:WKBp}) and (\ref{eq:WKBtau}) to give
\begin{align}
P_{\mathrm{eff}}&=\exp \left(-\frac{8}{3\hbar}w\sqrt{2mV_\mathrm{mf}} \left(1-\frac{b N^{1/3}}{V_{\mathrm{mf}}}\right)^{3/2} \right) \label{eq:WKBp_mf}\\
\tau_{\mathrm{eff}}&=\sqrt{2m}\frac{(x_0-w)}{\sqrt{b N^{1/3}}}+2\sqrt{2m}\frac{w \sqrt{N^{1/3}}}{V_{\mathrm{mf}}}, \label{eq:WKBtau_mf}
\end{align}
the modified JWKB tunneling probability and semi-classical oscillation time, which are subsequently substituted into Eqs.~(\ref{eq:Gamma_tri}) and (\ref{eq:decay_eq}).

\begin{figure}
 \subfloat[]{
 \includegraphics[width=8.3 cm]{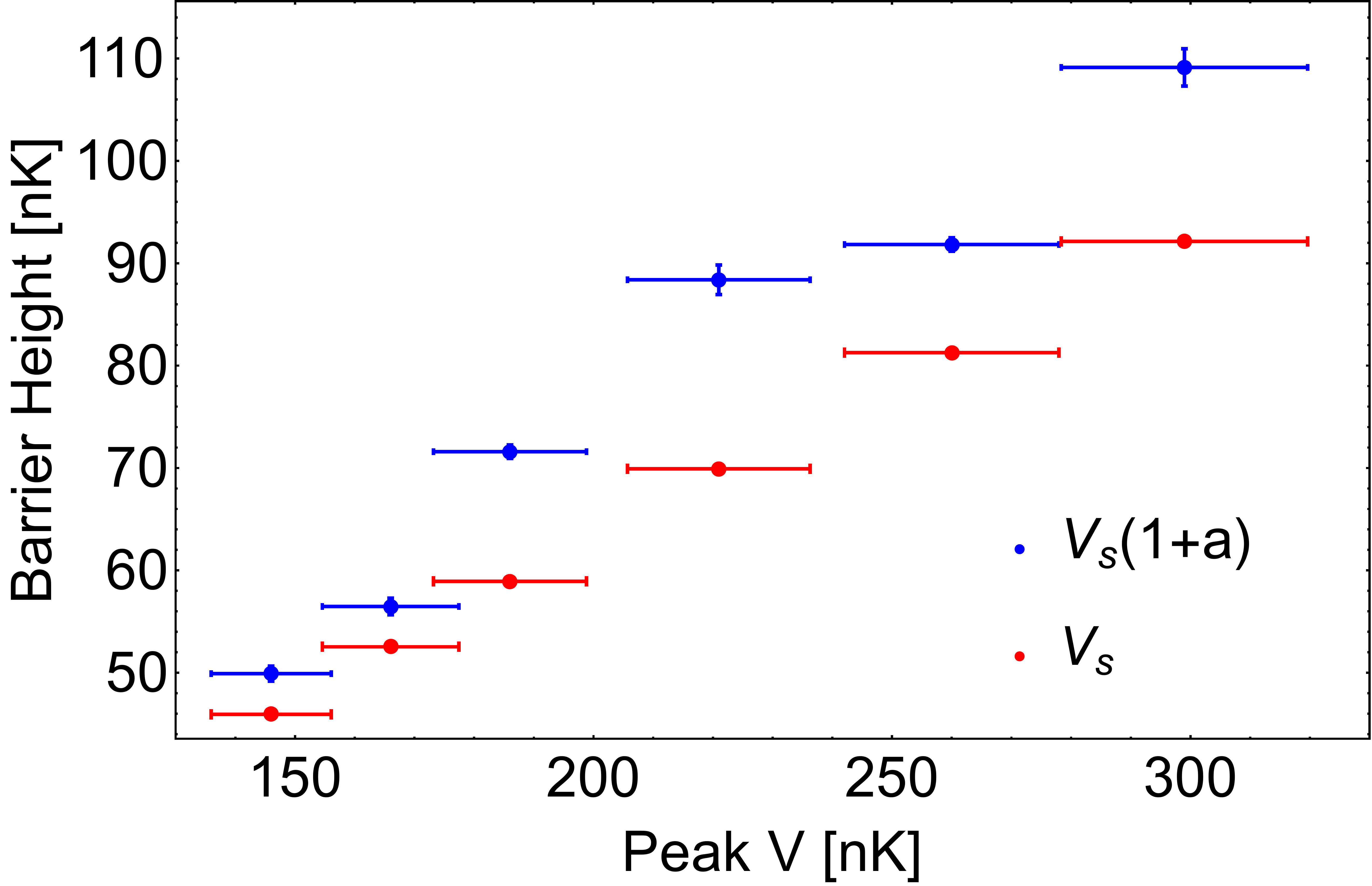}
\label{fig:WKBpara:a}
 }
 \hfill \\[-1ex]
 \subfloat[]{
 \includegraphics[width=8.3 cm]{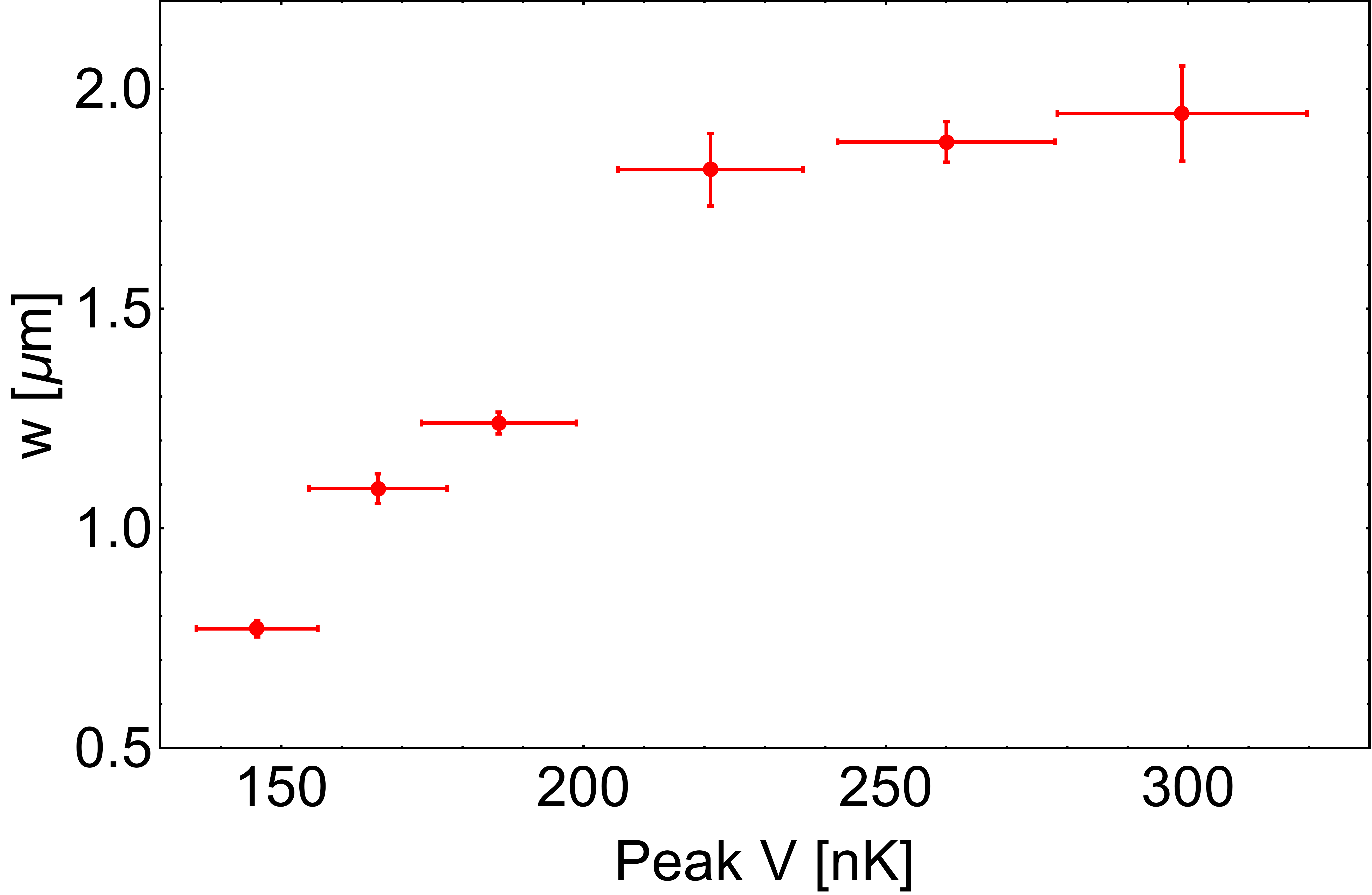}
 \label{fig:WKBpara:b}
 }
 \caption{\textit{Fitting Results in Quantum Tunneling Regime.} Optimized values of (a) effective saddle height (blue) at start of tunneling dynamics in comparison to the experimentally measured bare saddle height (red) vs. the bare peak height, showing how mean-field effects significantly correct the tunneling dynamics.  (b) Effective potential width $w$ for the weak configuration. Both the effective width and height increase as a function of bare experimental barrier height $V$.}
\label{fig:WKBpara}
\end{figure}

\begin{figure}
 \subfloat[]{
 \includegraphics[width=8.3 cm]{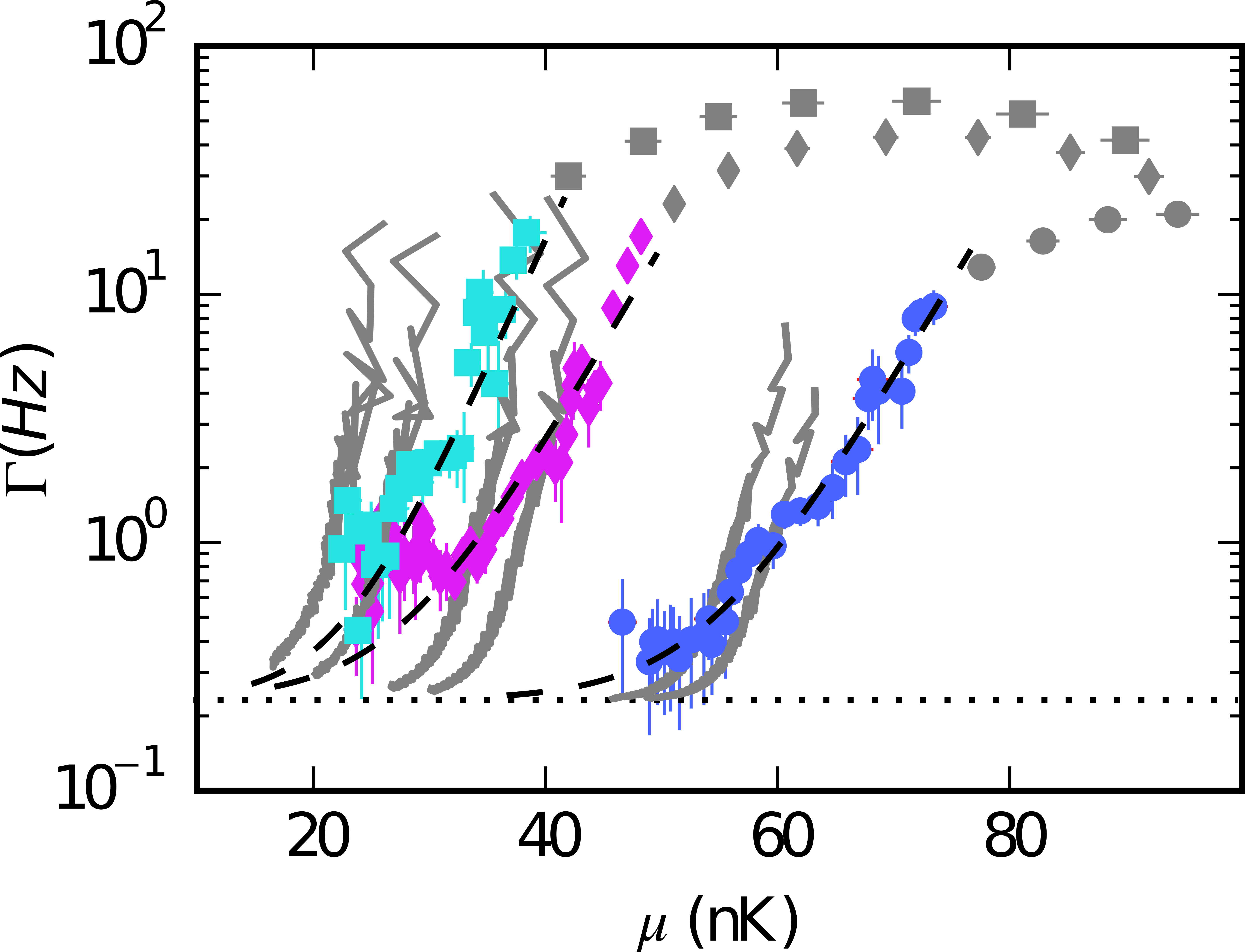}
\label{fig:loggammamu:a}
 }
 \hfill \\[-3ex]
 \subfloat[]{
 \includegraphics[width=8.3 cm]{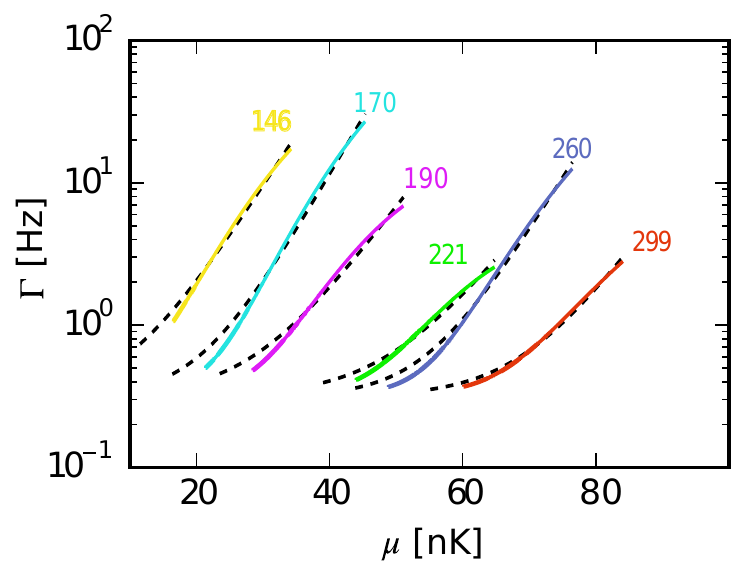}
 \label{fig:loggammamu:b}
 }
\caption{\textit{Experimental Data and Theoretical Curves Comparison: Decay Rate vs Chemical Potential.} (a) Experimental data for the weak trapping configuration (colored points), 3D GPE simulations (gray lines), and fit results to an exponential plus a constant, Eq.~\eqref{eq:decayrateexp} (dashed black lines). (b) Modified JWKB curves modeling this data (solid colored lines) and fit curves to our model, again based on Eq.~\eqref{eq:decayrateexp} (dashed black lines), showing that the basic exponential dependence on chemical potential is captured by a much simpler 1D modified JWKB approach. Numbers in (b) represent barrier height $V_{\rm s}$, and the same color scheme is used in (a).}
\label{fig:loggammamu}
\end{figure}

Figure~\ref{fig:WKBpara} contains the fitted potential FWHM parameter $w$, the saddle height $V_{\rm s}$, and the initial ($t=t_{\mathrm{QT}}$) mean-field saddle height $V_{\mathrm{mf}}=V_{\rm s}(1+a)$. The increasing trend in $w$ reflects the fact that the experimental trap widens with increasing height, as shown in Table~\ref{table:exptable}.
The trend in $V_{\mathrm{mf}}$ overall increases with increasing peak height, leading to a deeper potential well in Fig.~\subref*{fig:WKBpara:a}. Deeper wells are able to hold larger number of atoms, and the more atoms the stronger the mean-field effect (larger barrier). These larger barriers cause slower maximal tunneling rates as shown in Fig.~\subref*{fig:loggammamu:b}, except for 260\,nK. The case of peak height 260\,nK has a smaller increase from bare saddle height $V_{\rm s}$ to $V_{\mathrm{mf}}$ than expected; $V_{\mathrm{mf}}$ for 260\,nK is close to that of 221\,nK even though the bare height $V_{\rm s}$ is more than 10\% larger, indicating additional unaccounted for systematic error in the experimental setup for this barrier height.  The initial number of atoms for $V_{0}=260\,\mathrm{nK}$ ($N_{0}\approx 583,000$) is much closer to $V_{0}=190\,\mathrm{nK}$ ($N_{0}\approx560,000$) than $V_{0}=221\,\mathrm{nK}$ ($N_{0}\approx630,000$). This is further supported by the results in the tight configuration runs, see Section~\ref{section:2nd}.

In the experiment, we were able to fit a nearly exponential relation between the experimental tunneling rate $\Gamma_{\rm exp}$ and the chemical potential, Eq.~(\ref{eq:decayrateexp}), for weak (Fig.~\subref*{fig:loggammamu:a}) and tight (Fig.~\subref*{fig:Nt3D:c}) configurations.  Figure~\subref*{fig:loggammamu:b} shows both experiment and 3D mean-field simulations for the weak configuration with barrier heights of $V_{0}$ = 170(11)\,nK (teal squares), $V_{0}$ = 190(13)\,nK (purple diamonds) and $V_{0}$ = 260(18)\,nK (blue circles).  From the modified JWKB fits, we calculate the instantaneous decay rate $\Gamma$ as a function of chemical potential $\mu$, and use Eq.~(\ref{eq:decayrateexp}) to fit to this relation, Fig.~\subref*{fig:loggammamu:b}; in light of the discussion in Section~\ref{section:2nd}, we omit tight configuration data. By examining Fig.~\ref{fig:loggammamu}, one finds that the fit captures the gross features of both the experimental data and modified JWKB curves, especially when comparing curves which correspond with the same barrier height (teal, purple, and blue curves between both subfigures. In Fig.~\subref*{fig:loggammamu:b}, for barrier heights $V_{0} \leq 221\,\mathrm{nK}$, it can be seen that quantum tunneling is still contributing, as the tails for the fits strongly deviate from the modified JWKB curves.

\begin{figure}
\begin{center}
\includegraphics[width=8.3 cm]{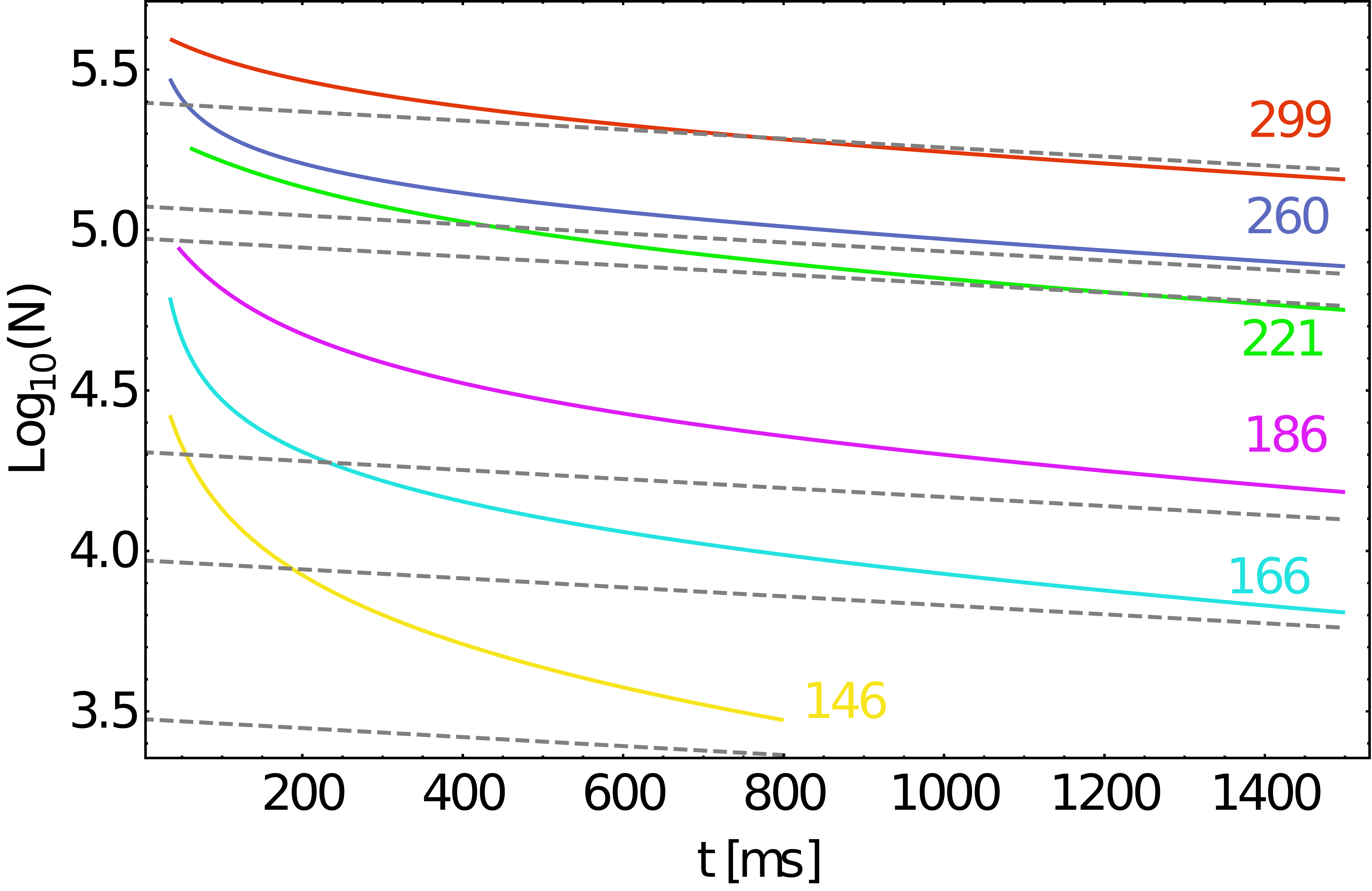}
\end{center}
\caption{\textit{Theoretical Fits of Trapped Particle Number for Weak Configurations.} All fits demonstrate same trends as the case study in Fig.~\ref{fig:fig186}. Gray dashed lines represent background loss for each barrier height. Colors indicate barrier heights for weak configuration, in nanokelvin. }
\label{fig:figcombine}
\end{figure}

Finally, we demonstrated the power of the resulting JWKB prediction for $N(t)$ in our case study in Fig.~\ref{fig:fig186}.  Figure \ref{fig:figcombine} shows $N(t)$ predictions from the theoretical fitting results from Fig.~\ref{fig:WKBpara} for the other five sets of data in the weak configuration, with the same trends of non-exponential MQT behavior.  For brevity, we do not illustrate complete error bar analyses here, but they show similar trends to the case study.

In Section.~\ref{section:case_study} we put forth the question of whether or not an effective 1D JWKB model could reproduce the experimental findings, and in this section we have demonstrated that it is indeed possible. Two underlying assumptions were used, that we could simplify to a 1D model and that the JWKB was applicable.  JWKB is applicable when the spatial derivative of the de Broglie wavelength $\lambda_{\mathrm{dB}}=\hbar/p(x)$, is small, $d\lambda_{\mathrm{dB}}/dx \ll 1$, where $p(x)$ is the semiclassical momentum.  Clearly this is the case here since our model fits the data well.  How can this be true for a 3D system?  It turns out that the 3D potential used here can be efficiently treated by averaging over all semi-classical paths, and using the JWKB approximation for tunneling through the barriers, as established in~\cite{makri_semiclassical_1989}. Our effective JWKB potential, Eq.~(\ref{eq:new_veff}), uses a flat potential inside the well, while a true 1D path would necessarily include the harmonic and Gaussian contributions from Eq.~(\ref{eq:trap}). However, this allowed the model to capture, on average, the classical oscillation frequency with the barriers caused by the complicated semi-classical trajectories in the true 3D well.  The key point is that our JWKB approach is an effective model, not a first principles calculation, and the results fit the data well.

\subsection{The Tight Configuration}\label{section:2nd}

\begin{figure}
\subfloat[]{
\includegraphics[width=8.3 cm]{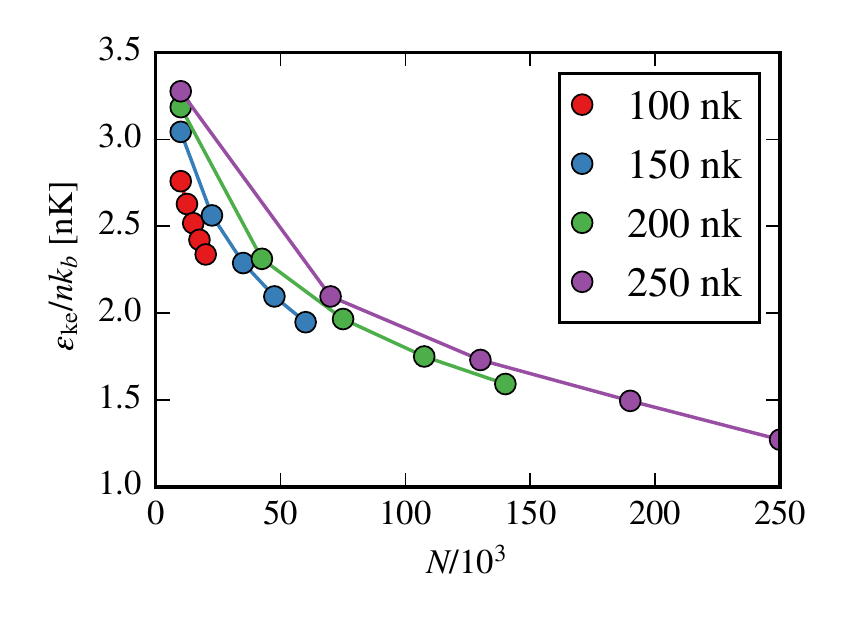}
}
\hfill \\[-2ex]
\subfloat[]{
\includegraphics[width=8.3 cm]{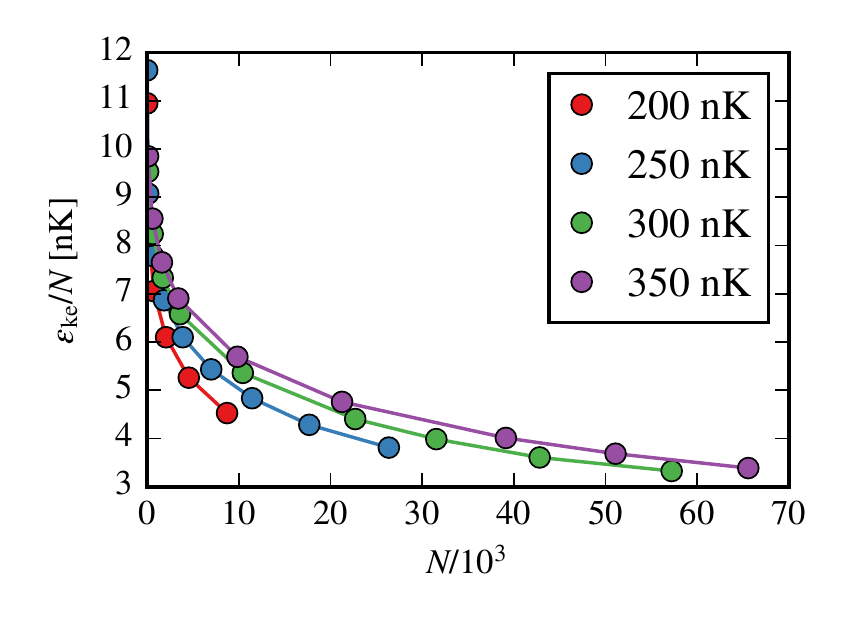}
}
\caption{\textit{Kinetic Energy Correction to the Chemical Potential.} In the (a) weak and (b) tight configurations, as the number of atoms increases, the kinetic energy contribution decreases, rendering the Thomas-Fermi approximation more accurate. Tight configuration requires kinetic energy a correction due to fewer particle numbers. Colored dots are from 3D GPE calculation while lines are a guide to the eye.}
\label{fig:mu2}
\end{figure}

According to our calculations from the 3D GPE simulations for the kinetic energy term, the chemical potential needs to be modified for the tight configuration potential. This is because we estimate the chemical potential with the Thomas-Fermi approximation~\cite{shreyas2016}. But, for the tight configuration, due to the tighter confinement, we have a smaller particle number in the potential, which is almost one-fifth of that in the weak configuration. This makes the kinetic energy term larger in the tight configuration where the acceleration is large, which leads to a modification of chemical potential of about 6 to 21\,nK for different barrier regimes in the tight configuration. Figure~\ref{fig:mu2} shows the kinetic energy term in the weak and tight configuration. These are calculated by numerically solving the full 3D GPE equation using imaginary time propagation. For the weak configuration, the kinetic term is much smaller, Thomas-Fermi is an excellent approximation and indeed it fits the data well. While on the contrary, the kinetic term becomes larger and can't be neglected in the tight configuration. All the plots which include the chemical potential in this Article are already fixed with its kinetic energy term according to Fig.~\ref{fig:mu2}.

The tight configuration has larger fluctuations and the traps deviate from their ideal shape in Eq.~(\ref{eq:trap}), an effect to which our theoretical model is sensitive; $V_{ts}$ in the tight configuration had about twice as much uncertainty as the weak configuration. Similar to the weak configuration, the largest contribution to the error envelopes was the uncertainty in barrier height and the fluctuation in the particle number, with all other errors typically 1-2 orders of magnitude smaller. Particle number error dominated early times ($t < 100$ ms) with errors up to $O(10^{3} \sim 10^{4})$ before dropping 1-2 orders of magnitude, and the barrier error dominated thereafter with errors up to $O(10^{3} \sim 10^{4})$. So, although the error contributions were not significantly larger in the tight configuration, due to the fewer total particles involved in tunneling these experimental uncertainties resulted in larger error envelopes relative to the number of particles.

\begin{figure}
 \subfloat[]{
 \includegraphics[width=8.3 cm]{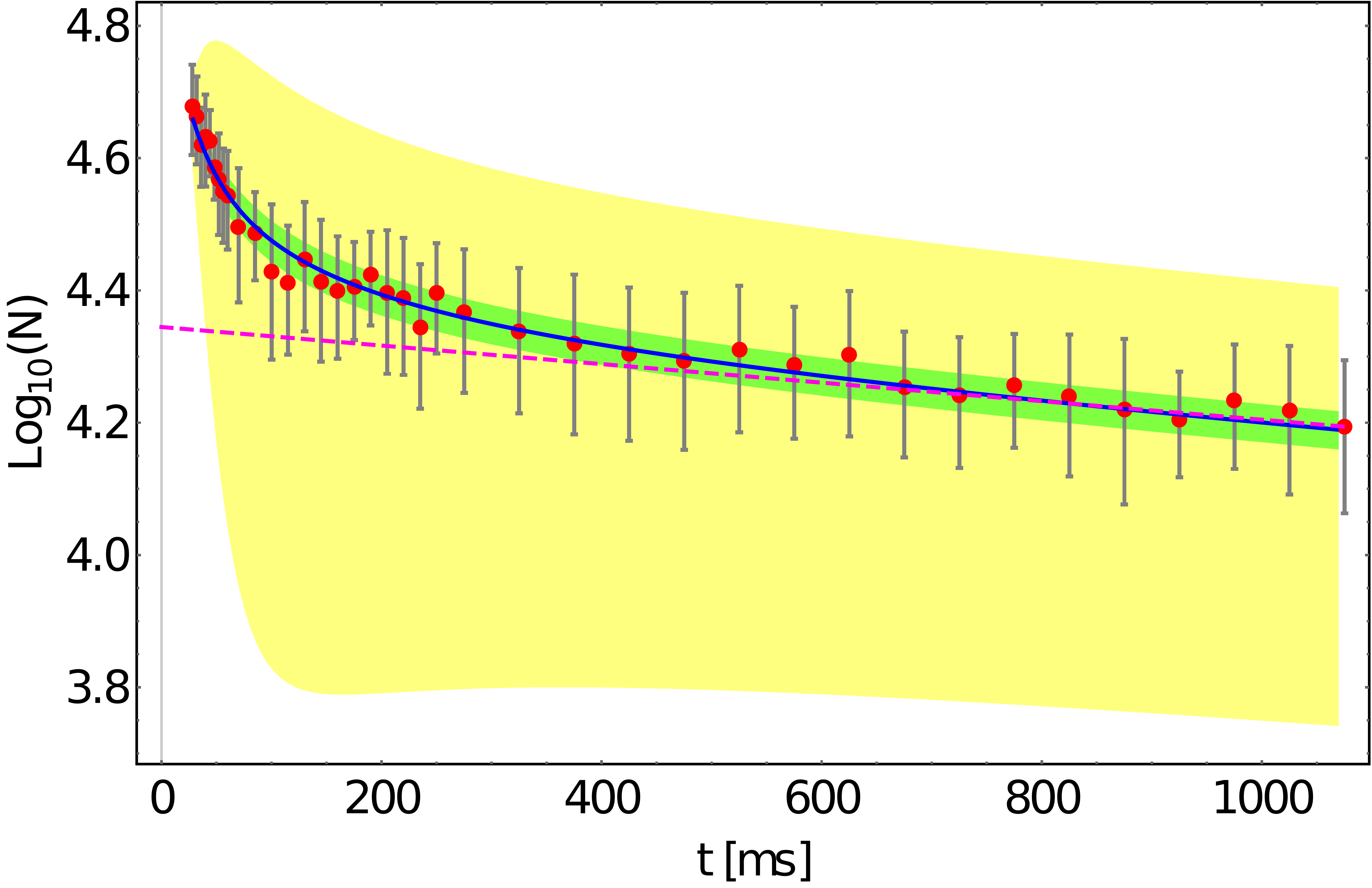}
 \label{fig:WKB2ndfit:a}
 }
 \hfill \\[-1ex]
 \subfloat[]{
 \includegraphics[width=8.3 cm]{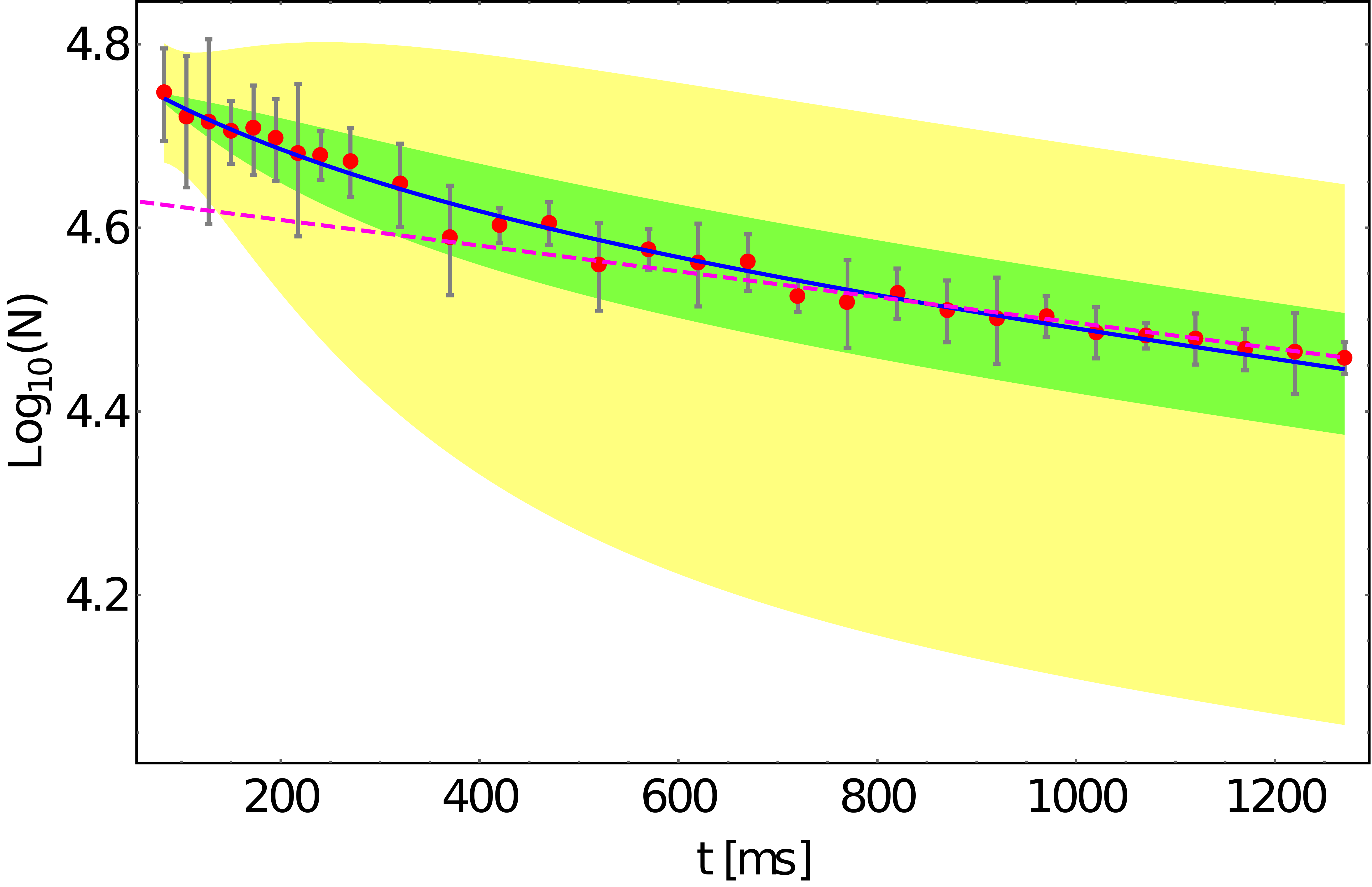}
 \label{fig:WKB2ndfit:b}
 }
 \caption{\textit{Experimental Data and Theoretical Fits in Tight Configurations.}  Experimental data for MQT (theoretical fits: blue lines) for the tight configuration shows either (a) for V=290\,nK, large experimental error and small fitting error, or (b) for V=330\,nK, few data points with large fitting error. Red points are the mean value of the number of atoms in the trap from experimental data, with $1\sigma$ error bars, and dashed pink line is background loss. The green regions indicate fitting error, and yellow regions combined uncertainty in: model fit, experimental parameters, and data points.}
\label{fig:WKB2ndfit}
\end{figure}

All fits for the tight configuration fall into two trends. Figure~\ref{fig:WKB2ndfit} shows two representative fits for peak heights of 290\,nK and 330\,nK. Both plots distinguish the error in the fit due to uncertainty in our fit parameters (green envelope), and the total uncertainty including experimental uncertainty (yellow envelope); contrast with weak configuration in Fig.~\ref{fig:fig186}, in which the total uncertainty is not much larger than the data error bars. Quantum tunneling is only observable for the first 600\,ms, with background losses quickly dominating the dynamics thereafter, in contrast to the weak configuration where MQT is observable for about twice as long.  Data sets for the tight configuration either had many points with large experimental error bars and large fluctuations like Fig.~\subref*{fig:WKB2ndfit:a}, or had few data points with smaller error bars but still large fluctuations like Fig.~\subref*{fig:WKB2ndfit:b}; all but $V_{0}=240\,\mathrm{nK}$ had reduced chi-squared values for $N(t)$ of $O(0.05\sim 0.10)$. Although the total uncertainty for the tight configuration did not allow for physical insight into the modified JWKB free parameters, the model was still able to capture the overall trend in Fig.~\ref{fig:WKB2ndfit}.

\section{Effective Mean-Field}\label{section:TEBD}

This section is outlined as follows. We first present the literature which motivates using a renormalized mean-field parameter for depleted and fragmented BECs. Next, we present the many-body Hamiltonian and a corresponding mean-field equation. Following that, we describe the method by which we create a meta-stable state to study quantum tunneling. We show the failure in a straightforward mean-field application, and how a modified interaction parameter captures the low-order many-body dynamics from TEBD. Finally, we draw conclusions from the MQT experiment, and suggest future studies to distinguish between true mean-field and many-body effects in MQT experiments.

Experimentally, BECs have overall been very well understood with mean-field theory in the form of the GPE. However, the presence of many-body processes can induce fluctuations, fragmentation, and depletion, thereby rendering mean-field models inaccurate or even ineffective.  Many techniques and approaches are used to move beyond the GPE~\cite{Proukakis2008}; of particular interest here are those using an effective interaction parameter~\cite{stoof_theory_1996,stoof_initial_1997,proukakis_microscopic_1998,shi_finite-temperature_1998,morgan_response_2004,morgan_quantitative_2005}.
We will demonstrate  how, similar to multi-component optical systems being well modeled by an effective scalar GPE-like equation derived from a multi-mode or large vector NLS, fragmented and depleted BECs described by a many-body Hamiltonian can, for the purposes of MQT, be well described by a mean-field model with an effective or renormalized interaction parameter.

Advances in nonlinear optics and strong connections to the NLS motivate the effective interaction parameter used in our JWKB model.
A deep theoretical understanding of the propagation and self-focusing of partially incoherent beams in nonlinear media, which can lead to spatial incoherent solitons, has been developed through several equivalent methods~\cite{christodoulides_equivalence_2001}: an infinite set of coupled nonlinear Schr\"odinger equations (coherent density approach)~\cite{christodoulides_theory_1997,christodoulides_incoherent_1997,coskun_dynamics_1998,eugenieva_elliptic_2000}, propagation equation for mutual coherence function~\cite{nayyar_propagation_1997,hasegawa_dynamics_1975,snyder_big_1998,shkunov_radiation_1998}, and self-consistent multimode theory~\cite{mitchell_theory_1997,christodoulides_multimode_1998,akhmediev_partially_1998,krolikowski_collision-induced_1999,sukhorukov_coherent_1999}.
Similarities in the propagation equation for the mutual coherence method to the NLS allow for analytical techniques to be extended for partially incoherent regimes; e.g., derivation of an analytical expression for the collapse threshold of spatially partially coherent beans in inertial bulk Kerr media~\cite{bang_collapse_1999}.
It has also been shown that BECs at finite temperature have analogous behavior to incoherent light in nonlinear media~\cite{buljan_incoherent_2005}, further suggesting that analogies to nonlinear optics can offer useful insight.

For the many-body dynamics, consider $N$ bosons at zero temperature in the canonical ensemble in a quasi-bound state, one suitable for quantum tunneling as laid out in this Article. One appropriate model to study many-body dynamics for this system is the Bose-Hubbard Hamiltonian (BHH), which can be invoked using an optical lattice of $L$ sites with deep enough sites for a tight binding and single band approximation, or alternately taking it as a discretization scheme in an appropriate limit:
\begin{equation}
\label{eq:BHH}
\hat{H} = -J\sum_{i=1}^{L-1}(\hat{b}_{i+1}^\dagger\hat{b}_i+\mathrm{h.c.})+\sum_{i=1}^L [\frac{U}{2}\hat{n}_i(\hat{n}_i-\hat{1})+V^{\mathrm{ext}}_i \hat{n}_{i}].
\end{equation}
In Eq.~(\ref{eq:BHH}), $U$ determines the on-site two-particle interactions and $J$ is the energy of hopping. An external box trapping potential, such as in schematic plot Fig.~\subref*{fig:schematic:a}, is given by $V_i^{\mathrm{ext}}$ with height $h$. The field operator $\hat{b}_{i}^\dagger$ ($\hat{b}_{i}$) creates (annihilates) a boson at the $i\mathrm{th}$ site, satisfying the usual commutation relation $[ \hat{b}_{i},\hat{b}_{j}^\dagger]=\delta_{ij}$, and  $\hat{n}_{i} \equiv \hat{b}_{i}^\dagger \hat{b}_{i}$. We will work in hopping units: energies are scaled to $J$ and time $t$ to $\hbar/J$.  To simulate the many-body dynamics of the BHH, we use TEBD, a matrix-product-state method which is able to efficiently simulate one dimensional many-body quantum systems, and allows access to a wide variety of many-body quantities like fluctuation and entanglement~\cite{openMPS,carr2012m,schollwock2005}.

To describe the system from a mean-field perspective, the discrete nonlinear Schr\"odinger equation (DNLS) may either be obtained via discretization of the NLS or from a mean-field approximation of the BHH~\cite{carr2009m}:
\begin{equation}
\label{eq:DNLS}
\textstyle i\hbar\dot{\psi_i}=-J(\psi_{i+1}+\psi_{i-1}) + g|\psi_i|^2 \psi_i+V^{\mathrm{ext}}_{i} \psi_i.
\end{equation}
In Eq.~(\ref{eq:DNLS}), the condensate order parameter, $\psi_i$, is normalized to the number of atoms, $N = \sum^{L}_{i=1} |\psi_i|^2$, and $g \equiv U$ from the BHH. Note however, $g$ in the DNLS is related, but not equivalent, to the interaction parameter in the GPE. To be exact, $g=g^{(1)}\int dx |W^{(0)}(x)|^{4}$, where $g^{(1)}$ is the quasi-1D interaction strength, proportional to scattering length, and $W^{(0)}(x)$ is the lowest order Wannier state; see~\cite{mishmashThesis} for details. Mean-field simulations are performed using a fourth-order Runge-Kutta adaptation of Eq.~(\ref{eq:DNLS}).  The BHH approaches the DNLS equation in the mean-field limit $N\to \infty$, $U \to 0$, $NU=\mathrm{const.}$, which we take in units of $J$, $NU/J=\mathrm{const.}$  Both the BHH and the DNLS are single band models, valid when the many-body wavefunction covers many sites and has variations larger than the lattice constant.

\begin{figure}
\subfloat[]{
 \includegraphics[width=8.3 cm]{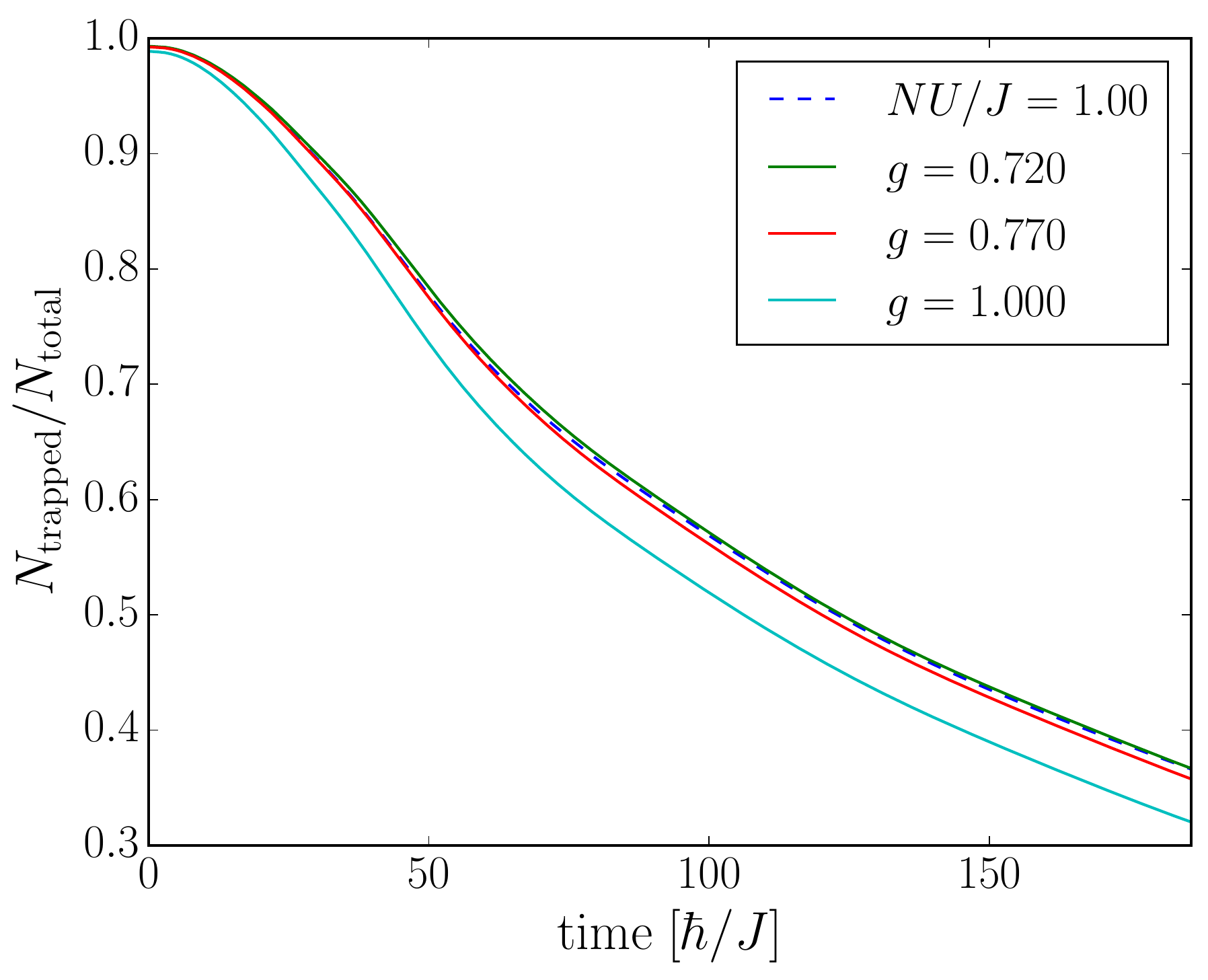}
 \label{fig:MF_Plot:a}
 }
 \hfill \\[-2ex]
 \subfloat[]{
 \includegraphics[width=8.3 cm]{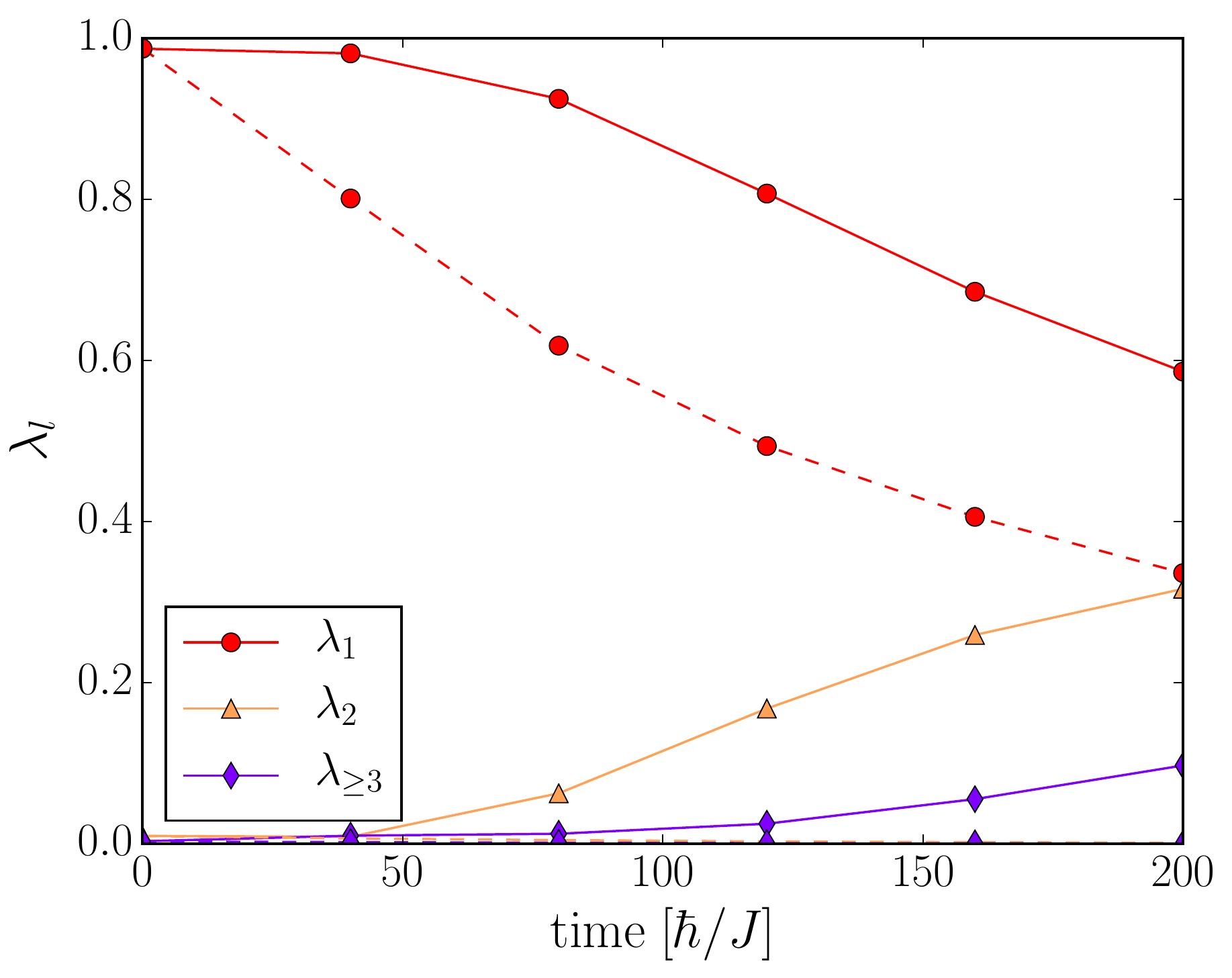}
 \label{fig:MF_Plot:b}
}
\caption{\textit{Effective Mean-Field, Many-Body, and Depletion} (a) Normalized trapped atoms for many-body ($NU/J=1.00$) is bounded from above and below by effective values $g=0.72$ and $g=0.77$ respectively, while direct mean-field comparison  $g=1.00$ under-predicts. (b) Semi-log two largest eigenvalues ($\lambda_{1}$,$\lambda_{2}$) and sum of remaining eigenvalues $\lambda_{\geq 3}$ of the single-particle density matrix for the total system (solid lines) and trapping well (dashed lines) are plotted versus time. The wavefunction has large occupation of two modes over the whole system (solid lines), with 10\% depletion as noted by non-zero $\lambda_{\geq 3}$. Points represent actual data with error bars smaller than marker, and lines are a guide to the eye.}
\label{fig:MF_Plot}
\end{figure}

We use box boundary conditions and initially set the barrier, $V^{\mathrm{ext}}_{i}$, to zero over the first 15 lattice sites, and $h=0.15$ over all the rest; the lattice size $L$ is large enough that the particles don't reach the end of the grid within simulation time, mimicking escape into open space. After propagation in imaginary time, the wavefunction is mostly trapped in the first 15 sites. We then reduce the external barrier to be $h=0.10$ from sites 16 to 19, thus rendering the state meta-stable, and evolve in real time. We reduce the height to induce larger fragmentation and depletion of the many-body wavefunction, but show how the renormalized mean-field still captures the key features of tunneling dynamics.

In Fig.~\subref*{fig:MF_Plot:a} we plot the normalized number of trapped atoms as a function of time for the many-body simulations with $NU/J=1.00$, mean-field with $g=NU/J=1.00$, and effective mean-field with two values of $g_{\mathrm{eff}}$. The values of $g_{\mathrm{eff}}$ were found by sweeping over $g$ values until appropriate upper and lower bounds on the many-body dynamics were found. In Fig.~\subref*{fig:MF_Plot:b}, we plot the two largest $\{\lambda_\ell\}$, the eigenvalues of the single particle density matrix $\langle \hat{b}_i^{\dagger}\hat{b}_j\rangle$, and $\lambda_{\geq 3}$, the sum of all but the two largest eigenvalues. We calculate eigenvalues for the single particle density matrix over the entire system ($\lambda_{l}^{\mathrm{sys}}$, solid curves in Fig.~\subref*{fig:MF_Plot:b}), as well as focusing in on the single-particle density matrix in the trapping potential ($\lambda_{l}^{\mathrm{trap}}$, dashed lines in Fig.~\subref*{fig:MF_Plot:b}). Information about the whole system $\lambda_{i}^{\mathrm{sys}}$ indicates a large degree of fragmentation, with up to 30\% occupation of the second mode, and depletion, with more than 10\% occupation in all but first 2 modes. Information regarding only the remaining atoms trapped in the well ($\lambda_{i}^{\mathrm{trap}}$) shows a large fraction of the BEC has escaped. Even though the many-body wavefunction would be considered over $30\%$ depleted by $t_{\mathrm{esc}}\approx 186$, the time at which the number of trapped atoms is $1/e$ of the initial value, our mean-field plots accurately bound the trend in many-body tunneling with a simple renormalization of the interaction parameter $g$ by about 25\%.
The effective mean-field simulation is able to qualitatively capture the overall trend in the number of trapped particles for tunneling, with less than $O(10^{-3})$ relative error in the trapped number of atoms, an error which would be indistinguishable within the error bars of many tunneling experiments including our own. While these results show that effective mean-field models can accurately reproduce low-order observables like trapped-atom density for depleted many-body wave-functions, we do not include loss, finite temperature, and other open-system effects in our many-body simulations.

In Section.~\ref{section:3Dsimu}, we show how a full 3D mean-field treatment of the experiment captures the gross features of the dynamics, and in Section.~\ref{section:WKB} we show how a JWKB model with an effective mean-field like parameter fits experiments runs well. These results are corroborated in this section, as we have shown how a renormalized mean-field interaction can adequately capture quantum tunneling of fragmented and depleted condensates, two effects that are likely present in an open quantum system such as our own, and compensated for with our JWKB parameter $a$. The applicability of mean-field seems to be larger than expected, and care must be taken to determine whether a given set of experimental data is purely mean-field.
Future experiments with sufficient resolution may be able to distinguish between many-body and effective mean-field dynamics by reducing number fluctuations in the initial state to resolve effective interaction strengths, for example by post-selecting on atomic number in measurements or using atom interferometry~\cite{Langen2015}. Furthermore,  coherence experiments on the escaped particles may further illuminate this distinction, as suggested by macroscopic occupation of more than one mode over the entire lattice in Fig.~\subref*{fig:MF_Plot:b}.

\section{Conclusions}
\label{section:Conclusions}

We first provided a brief discussion about different MQT regimes according to four factors: statistical properties and quasiparticles, the role of interactions, type of trapping potential, and dimension of the system.  In the discussion about atomic interactions, we suggested weak interactions can be described by mean-field-like theories, with fragmentation, entanglement and higher order quantum fluctuations increasing for stronger interactions and presenting experimentally untested regimes of MQT. We suggested different quasi-particle descriptions may result in distinct tunneling regimes.  We outlined how the shape of potential wells define the modes which particles can occupy, before and after tunneling, leading to a strong distinction between e.g. Josephson physics and quasibound many-body dynamics, or quantum escape.  We emphasize that the barrier is generally deformed by interactions and one must think at least in terms of an effective potential, as evidenced also by our experiment.  Although tunneling is primarily a 1D effect, higher dimensions can affect tunneling by creating chaos in semiclassical paths behind the barrier, for instance.  This brief survey set the tone for our own work and suggested future experiments and theoretical development needed, for instance, in the study of MQT of the unitary Fermi gas, perhaps in its holographic dual.

Then, we described the tunneling experiment and the resulting non-exponential decay of the trapped atoms. A mean-field description was validated by overall agreement between the experimental results and a 3D mean-field simulation, in decay curves and decay rates. The 3D mean-field simulation also calculated the kinetic term separately, which led to a correction of the kinetic term in the tight trap configuration. Our theoretical mean-field model reproduced the experimentally observed non-exponential decay, which was previously indicated to be the result of the participation of atomic interactions in the tunneling process. We then proceeded to explore the usefulness of a much simpler effective 1D model utilizing a modified version of the semiclassical approximation, or JWKB.
A case study of the experiment with barrier height 190nK in the weak configuration demonstrated the effectiveness of our 1D model. We further divided the tunneling process into three sub-regions: an initial decay region corresponding to classical spilling over the two saddles appearing in our potential; a non-exponential quantum tunneling regime during which interactions heavily affected the tunneling rate; and a background-loss dominated region. This pattern appeared in all the experimental runs.

Subsequently, we described the modified JWKB model in detail. The transition point into JWKB is when the relative height of the saddle points equals the chemical potential.
In the modified JWKB model we introduced an effective mean-field term which modified the barrier and produced non-exponential decay consistent with the experiment. Thus, this non-exponential decay can be described as generated by atomic interactions, which led to an effective dynamic barrier. Both 3D mean-field and 1D JWKB models confirmed an exponential relation between decay rate and chemical potential observed empirically in the experiments.  JWKB used only two fitting parameters, an effective mean-field saddle height $V_{\mathrm{mf}}$ resulting from unitless mean-field parameter $a$, and saddle width $w$. The parameter $w$ grew wider with increasing barrier height, which followed the experimental trend.  Increasing $V_{\mathrm{mf}}$ reflected the fact that higher peaks in the experiment have smaller maximum tunneling rates.

Finally, we showed how a renormalized mean-field theory is capable of capturing many-body quantum effects in low-order observables, by comparing discrete nonlinear Schr\"odinger and TEBD simulations, in analogy to optics contexts in which many modes create an effective scalar nonlinear-Schr\"odinger or Gross-Pitaevskii type description. Thus, experiments with large number fluctuations or error bars can have difficulty in discerning between mean-field and renormalized mean-field due actually to many-body effects. Future experiments which more precisely resolve atomic number and interaction strengths may be able to distinguish between the bare mean-field theories generally assumed to describe BEC dynamics and the effective or renormalized theories we have here suggested, in terms of a concrete observable, the number of atoms remaining in a many-body quasi-bound state as a function of time.

\begin{acknowledgments}
The authors thank Rockson Chang, Gavriil Shchedrin, and David Spierings for helpful discussions and A. Stummer for technical support. Computations
were performed on the gpc supercomputer at the SciNet HPC Consortium and Colorado School of Mines high performance computing resources as part of the Golden Energy Computing Organization. SciNet is funded by the Canada Foundation for Innovation under the auspices of Compute Canada, the Government of Ontario, Ontario Research Fund - Research Excellence, and the University of Toronto. This material is based in part upon work supported by NSERC, CIFAR, the US National Science Foundation under grant numbers PHY-1306638, PHY-1207881, and PHY-1520915 (LDC) , and the US Air Force Office of Scientific Research grant number FA9550-14-1-0287 (LDC), and China Scholarship Council (XXZ) and Northrop Grumman Aerospace Systems.
\end{acknowledgments}


%

\end{document}